\documentclass[onecolumn,preprintnumbers]{revtex4}
\usepackage{makeidx}
\usepackage{amssymb}
\usepackage{amsmath}
\usepackage{mathrsfs}
\usepackage{graphicx}
\usepackage{dcolumn}
\usepackage{bm}
\usepackage[center]{subfigure}
\usepackage{color}

\begin{document}

\title{A new form of liquid matter: quantum droplets}
\author{Zhihuan Luo$^1$, Wei Pang$^2$, Bin Liu$^3$}
\author{Yongyao Li$^{3}$}
\email{yongyaoli@gmail.com}
\author{Boris A. Malomed$^{4,5}$}
\affiliation{$^{1}$ Department of Applied Physics, South China Agricultural University,
Guangzhou 510642, China \\
$^{2}$ Department of Experiment Teaching, Guangdong University of
Technology, Guangzhou 510006, China \\
$^{3}$ School of Physics and Optoelectronic Engineering, Foshan University,
Foshan $528000$, China \\
$^{4}$ Department of Physical Electronics, School of Electrical Engineering,
Faculty of Engineering, and Center for Light-Matter Interaction, Tel Aviv
University, Tel Aviv 69978, Israel\\
$^{5}$Instituto de Alta Investigaci\'{o}n, Universidad de Tarapac\'{a},
Casilla 7D, Arica, Chile}

\begin{abstract}
This brief review summarizes recent theoretical and experimental results
which predict and establish the existence of \textit{quantum droplets}
(QDs), i.e., robust two- and three-dimensional (2D and 3D) self-trapped
states in Bose-Einstein condensates (BECs), which are stabilized by
effective self-repulsion induced by quantum fluctuations around the
mean-field (MF) states [alias the Lee-Huang-Yang (LHY) effect]. The basic
models are presented, taking special care of the dimension crossover, 2D $%
\rightarrow $ 3D. Recently reported experimental results, which exhibit
stable 3D and quasi-2D QDs in binary BECs, with the inter-component
attraction slightly exceeding the MF self-repulsion in each component, and
in single-component condensates of atoms carrying permanent magnetic
moments, are presented in some detail. The summary of theoretical results is
focused, chiefly, on 3D and quasi-2D QDs with embedded vorticity, as the
possibility to stabilize such states is a remarkable prediction. Stable
vortex states are presented both for QDs in free space, and for singular but
physically relevant 2D modes pulled to the center by the inverse-square
potential, with the quantum collapse suppressed by the LHY effect.

\vspace{6pt}
\textbf{\emph{Keywords}}: quantum droplet; Bose-Einstein condensate; Lee-Huang-Yang correction; votex state
\end{abstract}

\maketitle

\newpage
\tableofcontents
\newpage

\textbf{The list of acronyms} \smallskip

1D, 2D, 3D: one-, two, and and three-dimensional

3B: three-body (losses)

BEC: Bose-Einstein condensate

DDI: dipole-dipole interactions

FR: Feshbach resonance

GPE: Gross-Pitaevskii equation

GS: ground state

HO: harmonic-oscillator (potential)

LHY: Lee-Hung-Yang (corrections to the mean-field theory induced by quantum
fluctuations)

MF: mean field

NLS: nonlinear-Schr\"{o}dinger (equation or soliton)

QD: quantum droplet

TF: Thomas-Fermi (approximation)

TOF: time of flight

\section{Introduction}

The theoretical and experimental work with multidimensional (two- and
three-dimensional, 2D and 3D) solitons, i.e., self-trapped modes originating
from the balance between nonlinear self-attraction of wave fields and their
self-expansion driven, at the linear level, by diffraction and dispersion,
is more difficult in comparison to the commonly known concept of 1D solitons
\cite{r1,Malomed-2016-EPJST,Mihalache-2017-RRP,Kartashov-2019-NRP}. On the
one hand, 2D and 3D solitons offer many options inaccessible in 1D -- in
particular, the creation of 2D and 3D self-trapped states with the
topological charge, that represents intrinsic vorticity \cite%
{Malomed-2019-PD}, and more sophisticated 3D states, such as monopoles \cite%
{monopole}, skyrmions or hopfions with two independent topological
charges \cite{Volkov,skyrmion,Yasha}, and knots \cite{knots}. On the other hand, identification of
physically relevant models in which multidimensional solitons, both
structureless fundamental ones and higher-order states featuring topological
structures, are stable, and, eventually, creation of such objects in the
experiment, are challenging, because the most common cubic
self-attractive nonlinearity is only able to build \emph{completely unstable}
solitons in 2D and 3D, on the contrary to the ubiquitous one-dimensional
nonlinear-Schr\"{o}dinger (NLS) solitons, which are extremely robust
objects, realizing the ground state (GS) of the respective model \cite%
{Zakharov}. The fundamental cause of the instability is the fact that 2D and
3D NLS equations with the cubic self-attraction give rise, respectively, to the
critical and supercritical collapse, i.e., a trend to develop singular
solutions through catastrophic self-compression of the input \cite{Fibich}.
Therefore, an issue of profound interest is elaboration of physically
relevant settings which admit stabilization of 2D and 3D self-trapped
localized modes, fundamental and topologically structured ones alike \cite%
{r1,Malomed-2016-EPJST,Mihalache-2017-RRP,Kartashov-2019-NRP}.

Among several approaches to resolving this issue, an arguably most
successful one was theoretically proposed \cite{Petrov-PRL-2015} and
experimentally realized \cite%
{Cabrera-Sci-2018,Cheiney-PRL-2018,Semeghini-PRL-2018,Ferioli-PRL-2019,Errico-PRR-2019,FB-PRL-2016,Kadau-Nat-2016, FB-JPB-2016, Schmitt-Nature-2016, FB-2019-PRR, FB-2018-PRL, FB-2018-PRA, Chomaz-PRX-2016}
quite recently. It relies upon the effect of quantum fluctuations as a
correction to the mean-field (MF) dynamics of Bose-Einstein condensates
(BECs), which was first predicted long ago by Lee, Huang, and Yang (LHY)
\cite{LHY}, and was proposed to be used as a mechanism for the stabilization
of 3D self-trapped states in Ref. \cite{Petrov-PRL-2015}. In this context,
soliton-like states, which are called quantum droplets (QDs), form due to
the MF attraction, which may be provided either by contact (local)
inter-component attraction in binary BEC \cite{Petrov-PRL-2015}, \cite%
{Cabrera-Sci-2018,Cheiney-PRL-2018,Semeghini-PRL-2018,Ferioli-PRL-2019,Errico-PRR-2019}%
, or by long-range dipole-dipole interactions (DDIs) in a single-component
condensate of atoms carrying permanent magnetic moments \cite%
{FB-PRL-2016,Kadau-Nat-2016, FB-JPB-2016, Schmitt-Nature-2016, FB-2019-PRR, FB-2018-PRL, FB-2018-PRA, Chomaz-PRX-2016}. The collapse of the
droplets, which would take place in the MF approximation, is arrested by the
LHY effect, which is effectively represented by local quartic self-repulsive
terms in the respective NLS equations [that are usually called
Gross-Pitaevskii equations (GPEs), in the application to BEC]. The
competition between the MF attraction and LHY repulsion maintains a
superfluid\ state whose density (taking very low values) cannot exceed a
certain maximum, thus making it incompressible. This is a reason why this
quantum macroscopic state is identified as a fluid, and localized states
filled by it are called \textquotedblleft droplets". In addition to the 3D
QDs, droplets in the effectively 2D setting, which may exist under the
action of tight confinement in one direction, imposed by an external
potential, were theoretically predicted \cite{Petrov-PRL-2016} and
experimentally created \cite{Cabrera-Sci-2018,Cheiney-PRL-2018} too.

The objective of this article to provide a brief review of basic theoretical
and experimental results on the theme of QDs. First, basic models, in the
form of GPEs with the LHY corrections, are introduced in Section II, in the
full 3D form. Reductions of the equations to the 2D and 1D cases are also
presented.

Next, Sections III and IV provide a relatively detailed account of recent
experimental findings. These are stable QDs in a binary BEC composed of two
different hyperfine states of the same atomic species, as well as in a
heteronuclear mixture (Section III), and in single-component condensates of
magnetic-dipolar atoms (Section IV). In Section III, we also briefly address
collisions between 3D droplets in the binary condensate, which were
experimentally studied very recently \cite{Ferioli-PRL-2019}.

Sections V and VI report theoretical results. Because the current
theoretical literature on QDs is vast, while the size of this article is
limited, in these two sections we chiefly focus on most recent theoretical
predictions of stable three- and two-dimensional QDs with embedded
vorticity. Although they have not yet been reported in the experiment,
vortical QDs promise to realize a variety of novel features. Basic results
for stable \textquotedblleft swirling" (vortical) 3D and 2D droplets, with
unitary and multiple topological charges, are summarized in Section V. Also
included are results for vortex QDs in a semi-discrete 2D system \cite%
{Zhang-2019-PRL}. Another type of confined vortex modes is considered in
Section VI: effectively two-dimensional ones in the binary BEC pulled to the
center by the inverse-square potential, see Eq. (\ref{U}) below. In the
framework of the MF theory, all stationary states in this setting are
destroyed by the quantum collapse. However, the LHY terms suppresses the
collapse and help one to create an otherwise missing GS, as well as stable
vortex states \cite{Shamriz}. These solutions are singular at $r\rightarrow
0 $, but, nevertheless, physically relevant ones, as their norm (the number
of atoms in the condensate) converges.

Section VII concludes the article. In that section, we briefly mention
related topics which are not considered in detail in this review, and
discuss directions for the further development of studies in this area.

\section{Theoretical models of quantum droplets}

\subsection{Models of QDs in three, two, and one dimensions}

The energy density of a condensed Bose-Bose mixture in the MF\ approximation
is
\begin{equation}
\mathcal{E}_{\mathrm{MF}}=\frac{1}{2}g_{11}n_{1}^{2}+\frac{1}{2}%
g_{22}n_{2}^{2}+g_{12}n_{1}n_{2},  \label{MF}
\end{equation}%
where $g_{11}$, $g_{22}$, and $g_{12}$ are, respectively, the intra- and
inter-species coupling constants characterizing the interaction between
atoms, and $n_{j}$ is the density of the $j$-th component of the mixture. In
the case when intra-species interactions are repulsive, $g_{11,22}>0$, the
mixture is miscible, in the framework of the MF theory, if the intra-species
repulsion dominates over the inter-species interaction, $\sqrt{g_{11}g_{22}}%
>\left\vert g_{12}\right\vert $ \cite{Mineev}. Otherwise, the system is
immiscible (at $g_{12}>\sqrt{g_{11}g_{22}}$), or it collapses (at $g_{12}<-%
\sqrt{g_{11}g_{22}}$), if the attraction between the two species, accounted
for by $g_{12}<0$, is stronger than the effective single-species repulsion.

The celebrated LHY correction to the MF density (\ref{MF}), which is the
leading term in the beyond-MF energy density, originating from the
zero-point energy of the Bogoliubov excitations around the MF state \cite%
{LHY}, takes the following form, as derived by D. S. Petrov \cite%
{Petrov-PRL-2015}:
\begin{equation}
\mathcal{E}_{\mathrm{LHY}}=\frac{128}{30\sqrt{\pi }}gn^{2}\sqrt{na_{s}^{3}},
\label{5/2}
\end{equation}%
where $a_{s}$ is the \textit{s}-wave scattering length, $n$ is the density
of both components, assuming that they are equal, $g=4\pi \hbar ^{2}m/a_{s}$
is the corresponding coupling constant, and $m$ is the atomic mass. The LHY
term makes sense only for $a_{s}>0$, i.e., repulsive intra-species
interactions.

In this article, we concentrate on the consideration of symmetric modes in
the binary BEC, with equal components of the pseudo-spinor wave function.
Asymmetric states were considered too, in 1D \cite{Mithun}, 2D \cite%
{Li-PRA-2018}, and 3D \cite{Kartashov-PRA-2018} cases alike. The asymmetry
essentially affects stability of various modes. In particular, it tends to
strongly destabilize vortex modes with unequal topological charges in the
two components \cite{Kartashov-PRA-2018,Li-PRA-2018}. Further, QDs in
heteronuclear binary BEC\ \cite{Errico-PRR-2019} are always strongly
asymmetric, due to their nature.

In a dilute condensate, the LHY term, $\propto n^{5/2}$ in Eq. (\ref{5/2})
is, generally, much smaller than the MF ones, $\propto n^{2}$ in Eq. (\ref%
{MF}), hence the LHY correction is negligible. However, when the binary
condensate is close to the equilibrium point, at which the MF self-repulsion
in each component is nearly balanced by the attraction between the
components, the LHY term becomes essential or even dominant. The result is
the spontaneous formation of robust QDs, due to equilibrium between the
effective residual MF attraction (assuming, as said above, equal wave
functions of the two components) and the LHY repulsion, which stabilizes the
droplets against collapsing \cite{Petrov-PRL-2015}.

In this vein, by defining
\begin{equation}
\delta g\equiv g_{12}+\sqrt{g_{11}g_{22}},  \label{delta-g}
\end{equation}%
for $g_{11,22}>0$ and $g_{12}<0$, the latter condition corresponding to the
inter-species attraction, Ref. \cite{Petrov-PRL-2015} addressed the regime
with
\begin{equation}
0<-\delta g\ll g_{11,22}.  \label{<<}
\end{equation}%
The resulting LHY-amended GPE for the wave function of both components in
the symmetric 3D system can be written as
\begin{equation}
i\partial _{\tilde{t}}\phi =\left( -\frac{1}{2}\nabla _{\mathbf{\tilde{r}}%
}^{2}-3|\phi |^{2}+\frac{5}{2}|\phi |^{3}-\tilde{\mu}\right) \phi ,
\label{Petrov}
\end{equation}%
where $\mathbf{\tilde{r}}$, $\tilde{t}$, $\tilde{\mu}$ are, respectively,
rescaled coordinates, time, and chemical potential, as defined in Ref. \cite%
{Petrov-PRL-2015}, and the quartic self-repulsive term, $\left( 5/2\right)
|\phi |^{3}$, corresponds to the LHY energy density given by Eq. (\ref{5/2}).
The suppression of the collapse in Eq. (\ref{Petrov}) is guaranteed by
the fact that, for large values of the local density, $n$, the quartic
self-repulsion dominates over the cubic self-attraction, that drives the
onset of the collapse in the MF theory.

\begin{figure}[tbp]
\centering{\includegraphics[scale=0.6]{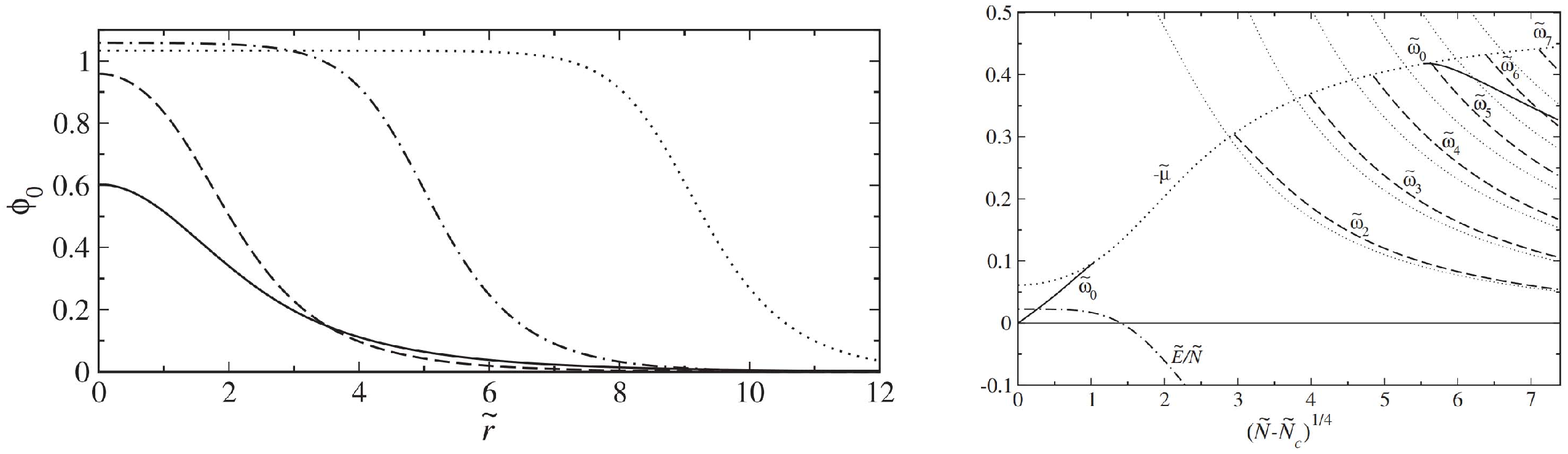}}
\caption{The left panel: radial profiles of isotropic 3D droplet's wave
function, versus the radial coordinate, for total norms $\tilde{N}%
=N_{c}\approx 18.65$ (solid), $\tilde{N}=30$ (dashed), $\tilde{N}=500$ (dash
dotted), and $\tilde{N}=3000$ (dotted). The right panel: the scaled energy
per particle $\tilde{E}/\tilde{N}$, defined as per Eq. (\protect\ref{E/N})
(the dash-dotted curve), the particle-emission threshold, $-\tilde{\protect%
\mu}$ (the thick dotted curve), the eigenfrequency of the monopole
excitation mode $\tilde{\protect\omega}_{0}$ (the solid line),
eigenfrequencies $\tilde{\protect\omega}_{l}$ of surface excitation modes
with the angular-momentum quantum number $l$ (dashed curves), and the
corresponding analytical approximation (thin dotted curves), versus $(\tilde{%
N}-\tilde{N}_{c})^{1/4}$. All the results are displayed as per Ref.
\protect\cite{Petrov-PRL-2015}.}
\label{fig-01}
\end{figure}

Equation (\ref{Petrov}) generates a family of stationary 3D-isotropic QDs,
which are looked for as%
\begin{equation}
\phi \left( \mathbf{\tilde{r}},\tilde{t}\right) =\phi _{0}\left( \tilde{r}%
\right) ,  \label{isotropic}
\end{equation}%
where $\tilde{r}$ is the radial coordinate. Radial profiles of the 3D
droplets, obtained as solutions of Eq. (\ref{Petrov}), as well as the
respective energy per particle,
\begin{equation}
\tilde{E}/\tilde{N}=\frac{1}{2}\left[ \int_{0}^{\infty }\phi _{0}^{2}(\tilde{%
r})\tilde{r}^{2}d\tilde{r}\right] ^{-1}\int_{0}^{\infty }\mathcal{E}%
(\tilde{r})\tilde{r}^{2}d\tilde{r}  \label{E/N}
\end{equation}%
(factor $1/2$ takes into account the contribution of both mutually symmetric
components into the total number of atoms), the particle emission threshold $%
-\tilde{\mu}$, and the spectrum of frequencies $\tilde{\omega}_{l}$ of the
droplet's surface modes, with angular-momentum quantum number $l$, are
displayed in Fig. \ref{fig-01}. One can find that, in a wide range of $(%
\tilde{N}-\tilde{N_{c}})^{1/4}$, where $\tilde{N}_{c}\approx 18.65$ is the
stability boundary (the QDs are unstable at $\tilde{N}<\tilde{N}_{c}$), all
excitation modes cross the threshold for sufficient small $\tilde{N}$, which
means that the modes, excited by initial perturbations on the surface of the
3D droplet, are depleted by emission of small-amplitude waves, in terms of
Eq. (\ref{Petrov}) \cite{Petrov-PRL-2015}.

Beyond-MF effects in lower dimensions, 2D and 1D, have also drawn much
interest. The respective models were derived in Ref. \cite{Petrov-PRL-2016},
starting from the full 3D setting and including tight transverse
confinement, imposed by an external potential acting in one direction, to
induce the 3D $\rightarrow $ 2D reduction, or in two directions, to impose
reduction 3D $\rightarrow $ 1D. In the 2D case, the energy density of the
symmetric pseudo-spinor condensate, with equal densities of the ``top" and
``bottom" spinor components, $n_{\uparrow \uparrow }=n_{\downarrow
\downarrow }\equiv n$, and equal scattering lengths, $a_{\uparrow \uparrow
}=a_{\downarrow \downarrow }\equiv a$, was derived in the form of
\begin{equation}
\mathcal{E}_{\mathrm{2D}}=\frac{8\pi n^{2}}{\ln ^{2}(a_{\uparrow \downarrow
}/a)}\left[ \ln (n/\left( n_{0}\right) _{\mathrm{2D}})-1\right] ,
\label{eq-Eng2D}
\end{equation}%
where $\left( n_{0}\right) _{\mathrm{2D}}=\left( 2\pi \right) ^{1}\exp
\left( -2\gamma -3/2\right) \left( aa_{\uparrow \downarrow }\right) ^{-1}\ln
(a_{\uparrow \downarrow }/a)$ is the equilibrium density of each component ($%
\gamma \approx 0.5772$ is the Euler's constant). The corresponding
LHY-amended GPE for the common wave function of both components reads
\begin{equation}
i\partial _{t}\psi =-\frac{1}{2}\nabla ^{2}\psi +\frac{8\pi }{\ln
^{2}(a_{\uparrow \downarrow }/a)}\ln \left( \frac{|\psi |^{2}}{\sqrt{e}%
\left( n_{0}\right) _{\mathrm{2D}}}\right) |\psi |^{2}\psi .  \label{log}
\end{equation}

The increase of the local density from small to large values leads to the
change of the sign of the logarithmic factor in Eq. (\ref{eq-Eng2D}). As a
result, the cubic term is self-focusing at small densities, initiating the
spontaneous formation of QDs, and defocusing at large densities, thus
arresting the transition to the collapse, and securing the stability of 2D
QDs.

\begin{figure}[tbp]
\centering{\includegraphics[scale=0.65]{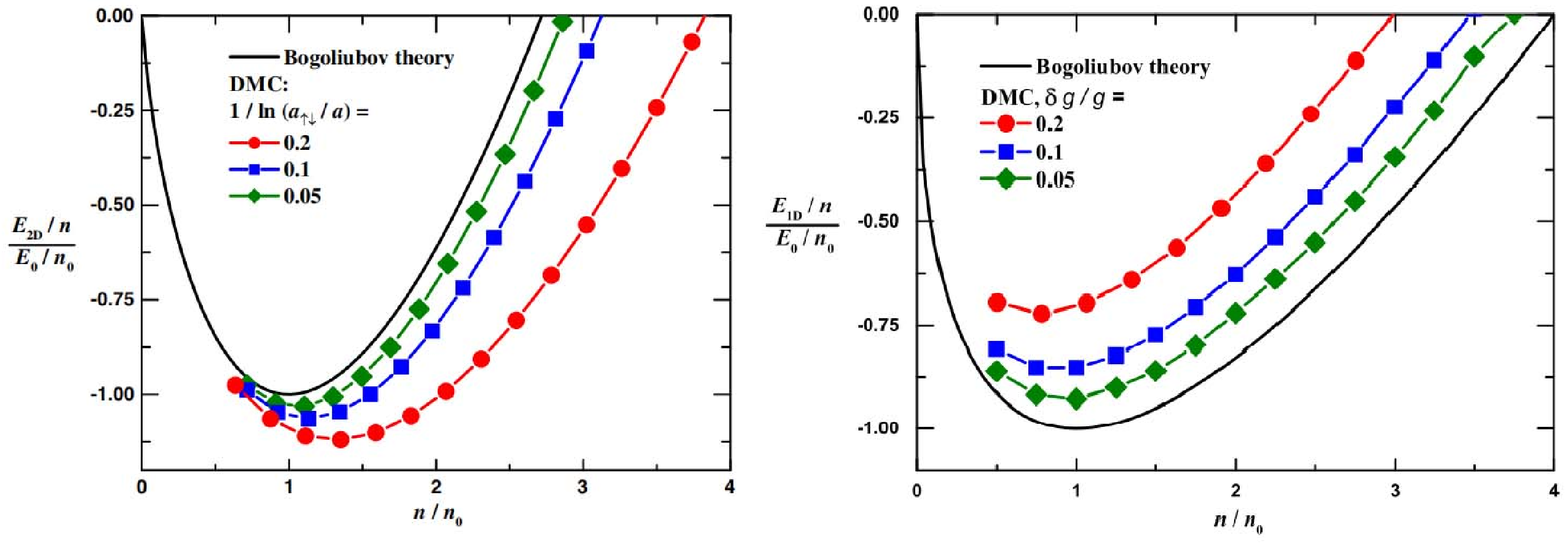}}
\caption{The energy per particle (rescaled) versus the atomic denisity, $n$
(rescaled), for the symmetric binary condensate. The left and rigth panels
present the results for the 2D and 1D settings, respectively. Solid lines
present predictions of the Bogoliubov approximation, as given by Eqs. (%
\protect\ref{eq-Eng2D}) and (\protect\ref{eq-Eng1D}), while lines with red
circles, blue squares, and green diamonds show data obtained for the
respective many-body settings, by means of the DMC method. The results are
displayed as per Ref. \protect\cite{Petrov-PRL-2016}.}
\label{fig-02}
\end{figure}

As shown in the left panel of Fig. \ref{fig-02}, the density dependence of
the energy per particle, calculated in the framework of the corresponding
many-body theory by means of the diffusion Monte Carlo (DMC) simulations
\cite{DMC}, converges toward the analytical result given by Eq. (\ref%
{eq-Eng2D}), with the decrease of $1/\ln (a_{\uparrow \downarrow }/a)$.

For the 1D setting, the analysis performed in Ref. \cite{Petrov-PRL-2016}
had yielded the following effective energy density:
\begin{equation}
\mathcal{E}_{\mathrm{1D}}=\delta g\cdot n^{2}-4\sqrt{2}(gn)^{3/2}/3\pi ,
\label{eq-Eng1D}
\end{equation}%
the corresponding equilibrium density being $\left( n_{0}\right) _{\mathrm{1D%
}}=8g^{3}/(9\pi ^{2}\delta g^{2})$. The respective LHY-amended GPE features
a combination of the usual MF cubic nonlinearity and a quadratic term,
representing the LHY corrections in the 1D geometry:
\begin{equation}
i\partial _{t}\psi =-(1/2)\partial _{xx}\psi +\delta g\cdot |\psi |^{2}\psi
-(\sqrt{2}/\pi )g^{3/2}|\psi |\psi .  \label{1D}
\end{equation}

Note that, in Eq. (\ref{1D}), LHY-induced quadratic term is self-focusing,
on the contrary to the defocusing sign of the quartic term in the 3D
equation (\ref{Petrov}). Because the most interesting results for QDs are
obtained in the case of the competition between the residual MF term and its
LHY-induced correction \cite{Astrakharchik-2018-PRA}, in the 1D case the
relevant situation is one with $\delta g>0$, when the residual MF
self-interaction is repulsive, in contrast with the residual self-attraction
adopted in the 3D setting, as mentioned above.

The energy per particle (\ref{eq-Eng1D}) is plotted, versus the rescaled
particle number, and for different values of $\delta g/g$, in right panel of
Fig. \ref{fig-02}. Similarly to the 2D case, for sufficiently small $\delta
g/g$, the scaled data produced by the DMC method for the many-body theory,
is in a good agreement with the Bogoliubov approximation given by Eq. (\ref%
{eq-Eng1D}).

\subsection{Dimensional crossover for quantum droplets}

\begin{figure}[tbp]
\centering{\includegraphics[scale=0.45]{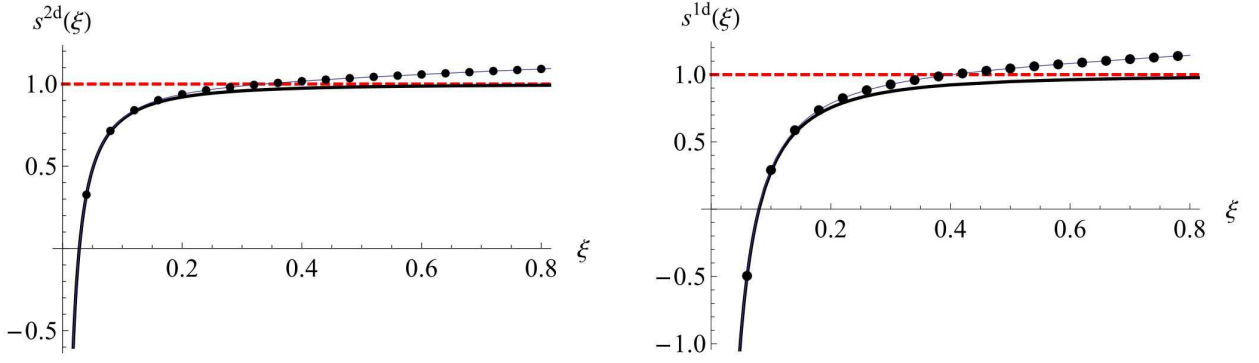}}
\caption{Ratios of the LHY energy densities at the $\mathrm{3D\rightarrow 2D}
$ and $\mathrm{3D\rightarrow 1D}$ crossovers, as functions of parameter $%
\protect\xi $, defined by Eq. (\protect\ref{xi}). The left panel: The thick
black line denotes the ratio computed numerically as per Eq. (\protect\ref%
{eq_LHY_2D_exact}), while the thin curve with dots represents the
approximate result given by Eq. (\protect\ref{eq_LHY_2D_approx}). The right
panel: The thick black line denotes the ratio computed numerically as per
Eq. (\protect\ref{eq_LHY_1D_exact}), while the thin curve with dots
represents the approximate result given by Eq. (\protect\ref%
{eq_LHY_1D_approx}). The results are presented as per Ref. \protect\cite%
{Zin-PRA-2018}.}
\label{fig-03}
\end{figure}
\begin{figure}[tbp]
\centering{\includegraphics[scale=0.65]{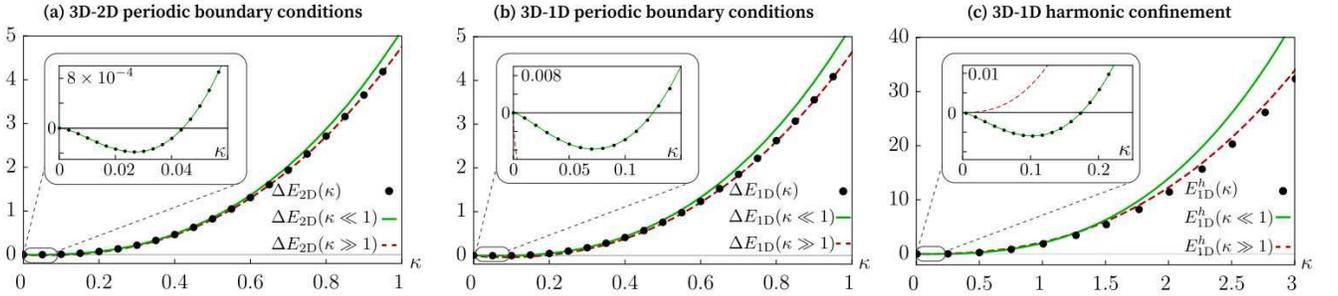}}
\caption{The crossover $\mathrm{3D\rightarrow 2D}$ and $\mathrm{%
3D\rightarrow 1D}$ is shown in terms of the beyond-MF corrections to energy
density, as functions of parameter $\protect\kappa $, see Eq. (\protect\ref%
{kappa}), pursuant to Ref. \protect\cite{Ilg-PRA-2018}. Black dots denote
the numerically exact resut. Analytically found asymptotic expressions for
small and large $\protect\kappa $ are plotted by the solid green and dashed
red lines, respectively. Panels (a) and (b) represent the $\mathrm{%
3D\rightarrow 2D}$ and $\mathrm{3D\rightarrow 1D}$ crossover, respectively,
with periodic boundary conditions in the transverse directions. (c) The $%
\mathrm{3D\rightarrow 1D}$ crossover with the transverse confinement imposed
by an harmonic-oscillator trapping potential.}
\label{fig-04}
\end{figure}

The quasi-2D and quasi-1D description of QDs outlined above is valid for
extremely strong transverse confinement. An estimate for experimentally
relevant parameters \cite%
{Cabrera-Sci-2018,Cheiney-PRL-2018,Semeghini-PRL-2018} yields a respective
estimate for the confinement size $L_{z}\ll l_{\mathrm{healing}}\sim 30$ nm,
where $l_{\mathrm{healing}}$ is the healing length in the condensate \cite%
{Shamriz}, while the actual value of $L_{z}$ used in the experiment is $\sim
0.6$ $\mathrm{\mu }$m. For this reason, the dimension crossover $\mathrm{%
3D\rightarrow 2D}$ requires a more careful consideration. In particular, for
a loosely confined (\textquotedblleft thick") quasi-2D layer of the
condensate it may be relevant to consider the 2D version of Eq. (\ref{Petrov}%
), keeping the quartic LHY term, which is, strictly speaking, relevant in
the full 3D space \cite{Shamriz}.

A detailed consideration of the dimensional crossover was presented in Ref.
\cite{Zin-PRA-2018}, which, similar to Ref. \cite{Petrov-PRL-2016},
addressed the binary BEC with repulsive intra-species interactions, $%
g_{11}\approx g_{22}>0$, and inter-species attraction, $g_{12}<0$. In
addition, periodic boundary conditions were imposed in the vertical
direction. As above, the system is tuned to be close to the balance
condition, defined as per Eqs. (\ref{delta-g}) and (\ref{<<}). In this case,
the effective quasi-2D energy density representing the LHY effect is defined
by the integration of the original 3D expression,
\begin{equation}
e_{\mathrm{LHY}}^{\mathrm{(2D)}}(\xi )=\lim_{r\rightarrow 0}\frac{\partial }{%
\partial r}\left( \frac{1}{2}r\sum_{q_{z}}\int d^{2}\mathbf{q}_{\perp }e^{i%
\mathbf{qr}}(\varepsilon _{\mathbf{q}}-A_{\mathbf{q}})\right) ,
\label{eq_LHY_2D_exact}
\end{equation}%
where $\varepsilon _{\mathbf{q}}=\sqrt{q^{4}+2\xi q^{2}}$, $A_{\mathbf{q}%
}=q^{2}+\xi $, the summation is performed with respect to discrete
wavenumbers in the vertical direction, while $\mathbf{q}_{\perp }$ is the 2D
wave vector in the horizontal plane. A crucially important parameter which
appears here is
\begin{equation}
\xi =(g_{11}n_{11}+g_{22}n_{22})/\varepsilon _{0},  \label{xi}
\end{equation}%
with $\varepsilon _{0}=(\hbar ^{2}/2m)(2\pi /L_{z})^{2}$. It determines the
ratio of the MF energy to the transverse-confinement energy.

For small $\xi \ll 1$, which is implied by the tight transverse confinement,
an approximate calculation of the energy density (\ref{eq_LHY_2D_exact})
yields
\begin{equation}
e_{\mathrm{LHY}}^{\mathrm{(2D)}}(\xi )=\frac{\pi }{4}\xi ^{2}\left( \ln (\xi
)+\ln (2\pi ^{2})+\frac{1}{2}+\frac{\pi ^{2}\xi }{3}\right) .
\label{eq_LHY_2D_approx}
\end{equation}

For the $\mathrm{3D\rightarrow 1D}$ crossover, one can define the effective
LHY energy density similarly, cf. Eq. (\ref{eq_LHY_2D_exact}):
\begin{equation}
e_{\mathrm{LHY}}^{\mathrm{(1D)}}(\xi )=\lim_{r\rightarrow 0}\frac{\partial }{%
\partial r}\left( \frac{1}{2}r\sum_{q_{x}q_{y}}\int dq_{z}e^{i\mathbf{qr}%
}(\varepsilon _{\mathbf{q}}-A_{\mathbf{q}})\right) .  \label{eq_LHY_1D_exact}
\end{equation}%
For small $\xi $, an approximate result is
\begin{equation}
e_{\mathrm{LHY}}^{\mathrm{(1D)}}(\xi )=-\frac{2\sqrt{2}}{3}\xi
^{3/2}+c_{2}\xi ^{2}+c_{3}\xi ^{3},  \label{eq_LHY_1D_approx}
\end{equation}%
where $c_{2}\simeq 3.06$ and $c_{3}\simeq 3.55$.

The dependence of ratios%
\begin{equation}
s^{\mathrm{(2D,1D)}}=e_{\mathrm{LHY}}^{\mathrm{(2D,1D)}}/e_{\mathrm{LHY}}^{%
\mathrm{(3D)}}  \label{s}
\end{equation}%
on $\xi $, where the 3D LHY energy density is $e_{\mathrm{LHY}}^{\mathrm{(3D)%
}}=16\sqrt{2}\pi \xi ^{5/2}/15$, are demonstrated in Fig. \ref{fig-03}. For
small values of $\xi $, the approximate expression matches the numerically
exact one well, while for large $\xi $, the ratios naturally approach $1$.

Another approach to calculating the beyond-MF corrections at the dimensional
crossover, which is based on the pioneering work \cite{Hugenholtz-PR-1959},
was elaborated in Ref. \cite{Ilg-PRA-2018}. In that work, a one-component
weakly interacting Bose gas satisfying the diluteness condition, $\sqrt{%
na_{s}^{3}}\ll 1$, is assumed to be confined in one or two directions by a
box potential with length $l_{\perp }$ and periodic boundary conditions. The
beyond-MF corrections to the energy density with the box potential, denoted
by $\Delta E_{\mathrm{2D,1D}}$, and, in addition, $\Delta E_{_{\mathrm{1D}%
}}^{h}$ for the $\mathrm{3D\rightarrow 1D}$ confinement imposed by the
harmonic-oscillator (HO) confinement (rather than by the box), which
represent the dimensionality reduction, are displayed, as functions of
\begin{equation}
\kappa \equiv na_{s}l_{\perp },  \label{kappa}
\end{equation}%
in Fig. \ref{fig-04}.

\section{Experimental observations of two-component quantum droplets (QDs)}

\subsection{Oblate (quasi-two-dimensional droplets)}

\begin{figure}[tbp]
\centering{\includegraphics[scale=0.5]{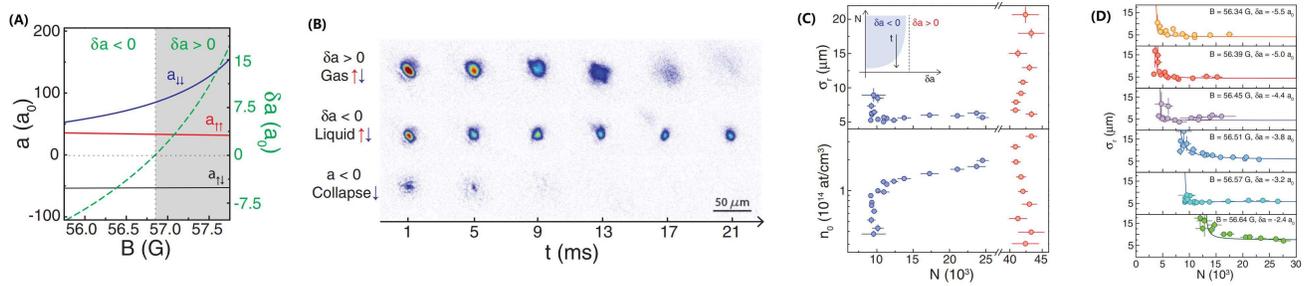}}
\caption{(A) Scattering lengths and the imbalance parameter, $\protect\delta %
a$ (see Eq. (\protect\ref{delta-a})) versus the magnetic field, $B$. (B) The
evolutions of in-situ images of the binary condensates at different times $t$%
. (Top) The expansion of a gaseous mixture, at $B=56.935$ G and $\protect%
\delta a=1.2a_{0}>0$. (Middle) The formation of a self-trapped droplet in
the binary condensate, at $B=56.574$ G and $\protect\delta a=-3.2a_{0}<0$.
(Bottom) The collapse of a single-component state $\left\vert \downarrow
\right\rangle $ in the attractive condensate, at $B=42.281$ G and $%
a=-2.06(2)a_{0}<0$. (C) The radial size of the mixture, $\protect\sigma _{r}
$ (top), and peak density, $n_{0}$ (bottom), as functions of the number of
atoms, $N$. (D) The dependence of $\protect\sigma _{r}$ on $N$ for different
magnetic fields $B$, from strong to weak attraction (top to bottom). These
experimental results are displayed as per Ref. \protect\cite%
{Cabrera-Sci-2018}.}
\label{fig-05}
\end{figure}

The creation of stable QDs in a BEC mixture of two Zeeman states of $^{39}$K
atoms, namely, $\left\vert \uparrow \right\rangle =\left\vert
F,m_{F}\right\rangle =\left\vert 1,-1\right\rangle $ and $\left\vert
\downarrow \right\rangle =\left\vert 1,0\right\rangle $, where $F$ is the
total angular momentum and $m_{F}$ is its projection was reported in Ref.
\cite{Cabrera-Sci-2018}. The potassium mixture is characterized by intra-
and inter-species scattering length $a_{\uparrow \uparrow }$, $a_{\downarrow
\downarrow }$, and $a_{\uparrow \downarrow }$. The residual MF interaction
is proportional to the effective scattering length
\begin{equation}
\delta a=a_{\uparrow \downarrow }+\sqrt{a_{\uparrow \uparrow }a_{\downarrow
\downarrow }}  \label{delta-a}
\end{equation}%
(cf. Eq. (\ref{delta-g})), which identifies the boundary between repulsive ($%
\delta a>0$) and attractive ($\delta a<0$) regimes. The interaction
strengths can be tuned, via the Feshbach resonance (FR) \cite{Feshbach}, by
an external magnetic field, $B$, as shown in Fig. \ref{fig-05}(A). The MF
energy of the mixture is proportional to $\delta a$, while the LHY
correction scales with the intra-species scattering lengths $a_{11}$, $%
a_{22} $. Atoms creating the QD are loaded in a plane of a vertical
blue-detuned lattice potential to compensate for gravity and a vertical
red-detuned optical dipole trap, which provides a horizontal radial
confinement. The experiment started with a sufficiently large magnetic
field, $B\approx 57.3$ G, which, via the Feshbach resonance, corresponds to $%
\delta a\approx 7a_{0}$, where $a_{0}$ is the Bohr radius. In this case, the
state of the binary BEC superfluid is miscible. Numbers of atoms in each
component can be measured by means of the Stern-Gerlach separation in the
course of free expansion of the gas, after the trapping potential was
switched off. Subsequently, the magnetic field is ramped down, to drive the
mixture into the attractive regime with $\delta a<0$, in which the radial
confinement is simultaneously switched off, letting the atoms move freely in
the horizontal plane. Panel (B) of Fig. \ref{fig-05} shows the evolution of
typical images at different moments of time $t$, following the removal of
the horizontal radial confinement, but keeping the vertical lattice
potential. In the case of $\delta a>0$, the mixture features a gas-like
expansion under an overall repulsive MF interaction, while for the
attractive regime (with $\delta a\approx -3.2a_{0}<0$), in which quantum
fluctuations, i.e., the LHY effect, start to dominate. As a result, a stable
self-trapped two-component droplet was observed in Ref. \cite%
{Cabrera-Sci-2018}. On the other hand, the same Fig. \ref{fig-05}(B)
demonstrates collapse occurring in a single-component condensate in the
attractive regime, dramatically different from the behavior of the binary
condensate.

Changes of radial size $\sigma _{r}$ and peak density $n_{0}$ of the
experimentally created QDs with the variation of the number of atoms, $N$,
are shown in panel (C) of Fig. \ref{fig-05}. In the attractive regime, both $%
\sigma _{r}$ and $n_{0}$ remain approximately constant at large $N$, as
expected for a liquid state. The existence of the mixture droplets require a
minimum number of atom, $N_{c}$, below which a liquid-to-gas transition
takes place, and the atomic cloud expands. Further, Fig. \ref{fig-05}(D)
shows strong dependence of the onset of the liquid-to-gas transition on
magnetic field $B$. The critical number $N_{c}$ increases with magnetic
field $B$, corresponding to the attenuation of the effective attraction in
the mixture.

A more accurate investigation of the liquid-to-gas transition of the binary
BEC was reported in Ref. \cite{Cheiney-PRL-2018}, by means of a similar
experiment in the mixture of two different atomic states in the potassium
condensate. The results also confirm the existence of QDs in the oblate
(quasi-2D) configuration. Moreover, it was found that traditional
matter-wave bright-soliton states, filled by the gaseous phase, and QDs,
filled by the ultradilute superfluid, coexist in a bistable regime,
providing an insight into the relation between these two kinds of
self-trapped states.

\subsection{Three-dimensional (isotropic) droplets}

\begin{figure}[tbp]
\centering{\includegraphics[scale=0.52]{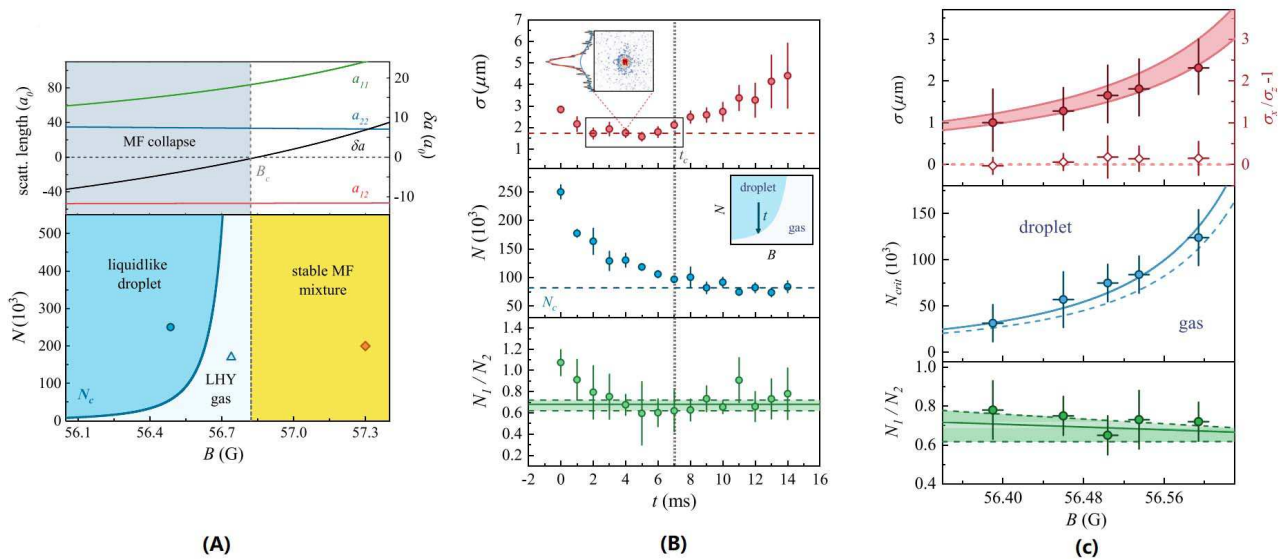}}
\caption{ (A) The dependence of the scattering lengths on the magnetic field
(top). The phase diagram of the binary condensate in the plane of the number
of atoms and magnetic field (bottom). (B) The time evolution of the
condensate's size $\protect\sigma $, the total atom number $N$, and the
population ratio, $N_{1}/N_{2}$ at $B=56.54$ G. (C) Measured values of $%
\protect\sigma $, $N_{c}$, and $N_{1}/N_{2}$ in established QDs,as functions
of the magnetic field, $B$. Lines and color stripes in (C) display
theoretical predictions for the QDs. The results are presented as per Ref.
\protect\cite{Semeghini-PRL-2018}.}
\label{fig-06}
\end{figure}

Following the original proposal by Petrov \cite{Petrov-PRL-2015}, QDs in the
full 3D space were created in a weakly interacting binary condensate of $%
^{39}$K \cite{Semeghini-PRL-2018}. The mixture is composed of two hyperfine
states of $^{39}$K. A cross dipole potential created by three red-detuned
laser beams, and an optical levitating potential were employed in the
experiment. The set of perpendicular beams was used to prepare the
condensate, while the later element helped to make the residual confinement
in all directions negligible. The residual scattering length of the mixture,
$\delta a=a_{12}+\sqrt{a_{11}a_{22}}$(cf. Eq. (\ref{delta-a})), decreases
with the external magnetic field, vanishing at $B_{c}=56.85$ G. As Fig. \ref%
{fig-06}(A) shows, at $B<B_{c}$, i.e., in the case of $\delta a<0$, the
binary condensate may be either a QD or an LHY gas, the two phases being
separated by a critical number of atoms, $N_{c}$. When released from the
external dipole trap in the attractive regime, the condensate with $N>N_{c}$
keeps a constant size, i.e., it demonstrates a well-defined QD, while, in
the case of $N<N_{c}$ it expands like a gas, as might be expected. In Fig. %
\ref{fig-06}(A), the average size of the atomic cloud is $\sigma =\sqrt{%
\sigma _{x}\sigma _{z}}$, where $\sigma _{x,z}$ are half-widths of the
density profile in the $x$ and $z$ directions, at the level of $1/\sqrt{e}$.

The evolution of the condensate's size $\sigma $ in the QD phase at $%
B=56.54\ $G, as well as the total atom number, $N$, and the population ratio
of two atomic states in the mixture, $N_{1}/N_{2}$, are presented in Fig. %
\ref{fig-06} (B). As seen in middle panel, $N$ rapidly drops during a few
milliseconds, because of losses induced by three-body (3B) inelastic
collisions, and eventually attains the critical values, $N_{c}$, at $t_{c}=7$
ms, where the liquid-to-gas phase transition takes place. The size, $\sigma
(t)$, exhibits a nearly constant value within the time interval $2$ ms $<t<$
$t_{c}$, as seen in the top panel, which confirms the establishment of a QD.
Afterwards, it expands as gas. In the course of the liquid-to-gas
transition, the total number of atoms, $N$, and the population ratio, $%
N_{1}/N_{2}$, remain constants.

The measurements for $\sigma $, $N$, and $N_{1}/N_{2}$ at critical point $%
N_{c}$ were extended to different values of magnetic field $B$, as shown in
Fig. \ref{fig-06}(C). The droplet's size $\sigma $ and critical atom number $%
N_{c}$ increase with the increase of the magnetic field. The colored area in
top panel corresponds to the theoretical prediction for $\sigma $ in the
range of norms $N_{c} \leq N \leq 2N_{c}$. The anisotropy measure, $\sigma
_{x}/\sigma _{z}-1$, remains zero for different magnitudes of $B$, revealing
that the droplet is a spherical isotropic one. The dependence of the
critical atom number, $N_{c}$, on $B$, shows good agreement with values
theoretically in Ref. \cite{Petrov-PRL-2015} for the metastable and stable
(dashed and solid lines, respectively) self-trapped states solution.

\subsection{Collisions between quantum droplets}

\begin{figure}[tbp]
\centering{\includegraphics[scale=0.75]{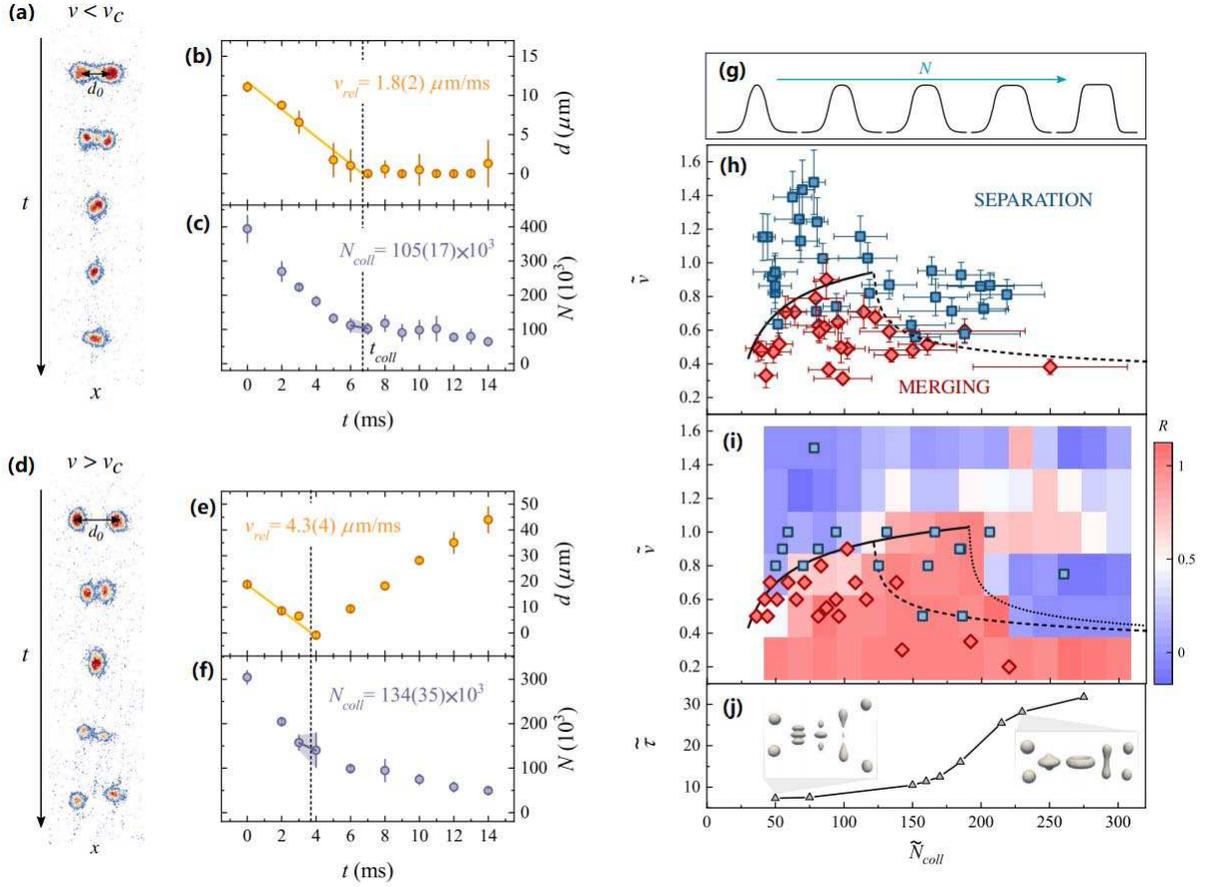}}
\caption{Typical examples of the observation of collisions between identical
droplets, resulting in the merger (a)-(c) and separation (passage) (d)-(f)
of the coliding droplets. Panels (a) and (d) show density profiles of the
colliding pair at different times. Panels (b) and (e) show the corresponding
evolution of distance $d$ between the droplets. Panels (c) and (f) display
the corresponding evolution of the total number of atoms, $N$. (g) The
droplet's wave functions corresponding to increasing values of $\tilde{N}$.
(h) Outcomes of the collision observation as a function of $\tilde{v}$ and $%
\tilde{N}_{\mathrm{coll}}$. (i) Results of numerical simulations, as a
function of $\tilde{v}$ and $\tilde{N}_{\mathrm{coll}}$. In (i), data points
and the color plot of ratio $R_{n}=n_{\mathrm{cm}}/(n_{\mathrm{cm}}+n_{%
\mathrm{out}})$, where $n_{\mathrm{cm}}$ is the density at the center of
mass and $n_{\mathrm{out}}$ is the peak density of the outgoing clouds,
represent results of the numerical calculation with and without 3B
(three-body) losses, respectively. The solid lines in (h) and (i) stand for
the asymptotic form of $\tilde{v}_{c}$ at small $\tilde{N}$, $\tilde{v}%
_{c}\propto \protect\sqrt{2|\tilde{E}_{\mathrm{drop}}|/\tilde{N}}$, while
the dotted line in (i) is $\tilde{v}_{c}\propto (\tilde{N}-\tilde{N_{0}}%
)^{-1/6}$, which is the predicted as the asymptotic form valid at large $%
\tilde{N}$. The dashed lines in (h) and (i) correspond to the same $\tilde{N}%
^{-1/6}$ scale, used just as a guide to the eye. (j) The timescale of the
collision, $\tilde{\protect\tau}$, as a function of $\tilde{N}$. Two insets
display examples of the collisional dynamics produced by the simulations
without three-body losses, for the two opposite cases of small and large $%
\tilde{N}$. The results are presented as per Ref. \protect\cite%
{Ferioli-PRL-2019}. }
\label{fig-07}
\end{figure}

Collision of moving classical droplets may lead to their merger into a
single one, provided that the surface tension is sufficient to absorb the
kinetic energy of the colliding pair. Otherwise, the colliding droplets
separate into two or more ones after the collision \cite{droplet}. Similar
phenomena in collisions of two component QDs were recently experimentally
demonstrated in Ref. \cite{Ferioli-PRL-2019}. That work exhibits two
different outcomes of the collision, i.e., merger and separation (passage).

For given magnetic field, there exists a critical velocity $v_{c}$, such
that the colliding QDs merge at $v<v_{c}$, and separate at $v>v_{c}$.
Typical examples of the evolution of the colliding droplets are displayed in
Fig. \ref{fig-07}. In the case of $v<v_{c}$, as shown in panel (a), the
distance between the droplets decreases and finally stays being equal to
zero, which implies the merger, as depicted in panel (b). By contrast, if
the case of $v>v_{c}$, the kinetic energy of the moving QDs overcomes the
surface tension, driving the separation after the collision, as shown in
panel (d). As a consequence, the distance between the separating droplets
increases, see panel (e). Further, it is shown in panels (c) and (f) that,
in both cases, the total atom number decreases due to the strong 3B loss in
the system \cite{Semeghini-PRL-2018}.

Figure \ref{fig-07}(h) presents a summary of results of experimentally
observed collision in the plane of the rescaled atom number $\tilde{N}$ and
velocity $\tilde{v}$, as produced in Ref. \cite{Ferioli-PRL-2019}. The
critical velocity $\tilde{v}_{c}$, which is the boundary between the merger
(red diamonds) and passage (blue squares), exhibits different dependences on
the number of atoms at small and large $\tilde{N}$, due to different energy
scales dominating in these cases. In the regime of incompressibility at
large $\tilde{N}$, the surface energy dominates, while the bulk and gradient
energies may be negligible. Therefore, $v_{c}$ is proportional to $\tilde{N}%
^{-1/6}$. In the opposite case of small $\tilde{N}$, the bulk energy has to
be taken into account, because it cannot be separated from the surface
energy. In this case, the consideration of the energy balance yields $\tilde{%
v}_{c}\propto \sqrt{2|\tilde{E}_{\mathrm{drop}}|/\tilde{N}}$. The
corresponding numerical results for the droplet collision, both with- and
without the 3B loss, is displayed in Fig. \ref{fig-07}(i). It is seen that
the numerical simulation with 3B loss is in good agreement with the
experimental results.

To probe the timescale of the collisions for various $\tilde{N}$, the
velocity of moving droplets, $v$, is set to be slightly larger than $\tilde{v%
}_{c}$, to ensure that the separation takes place after the collision. The
dependence of the timescale of the collision, $\tilde{\tau}$, on $\tilde{N}$
is displayed in panel (j), revealing that, in the liquid-like regime at
large $\tilde{N}$, the separation corresponds to longer timescales, because
in this limit the colliding pair forms a single cloud in an excited state
for a certain time interval, and they separate afterwards. Results of the
corresponding numerical simulation, without the 3B loss, are shown in insets
of Fig. \ref{fig-07}(j).

\subsection{Droplets in a heteronuclear bosonic mixture}

As outlined above, QDs were first created in mixtures of two different spin
states of $^{39}$K atoms. The mechanism stabilizing two-component QDs
applies as well to mixtures of different atomic species. Experimentally,
this possibility was realized in Refs. \cite{Errico-PRR-2019} and \cite%
{hetero2}, using a binary condensate of $^{41}$K and $^{87}$Rb atoms. The
results confirm the existence of stable droplets in the regime of relatively
strong inter-species attraction, and expansion of the mixture in the case
when the attraction is too weak. In particular, stable QDs were observed
with the ratio of atom numbers $N_{\mathrm{K}}/N_{\mathrm{Rb}}\approx 0.8$,
which is consistent with the results predicted by means of the analysis
based on Ref. \cite{Petrov-PRL-2015}. The heteronuclear droplets were
demonstrated to have a lifetime $\sim 10$ ms, much longer than ones created
in the binary condensate of $^{39}$K, that, as mentioned above, was
determined by 3B losses. The substantially longer lifetime offers one an
opportunity to gain insight in intrinsic properties of the QDs, such as the
observation of self-evaporation.

\section{Single-component QDs in dipolar condensates}

The theoretical and experimental findings summarized above demonstrate the
possibility of the creation of stable droplets in binary BECs, based on the
competition of the cubic nearly-balanced attraction between the two
components and self-repulsion in each of them, and the additional quartic
LHY-induced self-repulsion, see Eq. (\ref{Petrov}). Still earlier
experiments had produced robust QDs in single-component dipolar BECs
made of dysprosium \cite{FB-PRL-2016,Kadau-Nat-2016, FB-JPB-2016, Schmitt-Nature-2016, FB-2019-PRR, FB-2018-PRL, FB-2018-PRA} and erbium \cite{Chomaz-PRX-2016} atoms. Generating droplets in this setting is possible with the
attraction provided by the long-range dipole-dipole interaction (DDI), and
the stabilizing repulsion induced by the contact interaction, including the
LHY term. The dipolar BEC are characterized by the scattering length $a_{s}$
of the contact interaction, and the effective DDI length, $a_{dd}$.
Accordingly, the interplay between the DDI and the contact interactions is
controlled by parameter
\begin{equation}
\varepsilon _{dd}=a_{dd}/a_{s}.  \label{dd}
\end{equation}

\subsection{Quantum droplets in the condensate of dysprosium}

In the experiments reported in Refs. \cite{FB-PRL-2016,Kadau-Nat-2016},
isotope $^{164}$Dy with dipolar length $a_{dd}\simeq 131a_{0}$, where $a_{0}$
is the Bohr radius, is employed to create BEC. The background scattering
length of the contact interactions is $a_{\mathrm{bg}}=92a_{0}$, which was
modulated by many FRs. When $\varepsilon _{dd}$ is close to $1$ (see Eq. (%
\ref{dd}), the MF contact interactions and DDI nearly balance each other,
making the contribution from the beyond-MF\ LHY effect crucially important
for the creation of QDs.

\begin{figure}[tbp]
\centering{\includegraphics[scale=0.7]{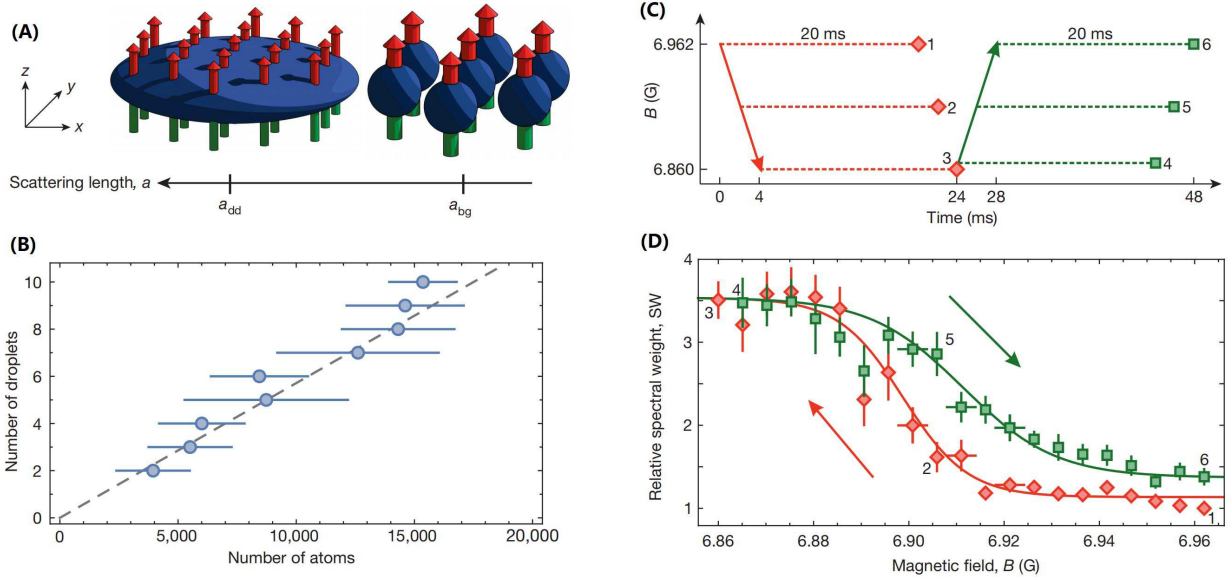}}
\caption{ A) A stable dipolar condensate of dysprosium atoms, with $%
a_{s}\approx a_{dd}$, loaded in a pancake-shaped trap (left). By reducing
scattering length $a_{s}$ to values close to $a_{\mathrm{bg}}$, atoms
coalesce into droplets, which build a triangular pattern (right). (B) The
number of droplets, $N_{d}$, as a function of the mean number of atoms in
the condensate. (C) The dysprosium BEC was prepared at magnetic field $%
B=6.962$ G, which was subsequently ramped down to $6.860$ G at a constant
change rate. After a waiting time of $20$ ms, the magnetic field was
increased, at the same ramp speed, back to higher values. (D) The hysteresis
plot of the spectral weight $\mathrm{SW}$ (see Eq. (\protect\ref{SW})) for
the structured patterns. All results are presented as per Ref. \protect\cite%
{Kadau-Nat-2016}.}
\label{fig-08}
\end{figure}

In Ref. \cite{Kadau-Nat-2016}, a stable BEC containing $\sim 10^{4}$
dysprosium atoms was created, by tuning the magnetic field to $B_{\mathrm{BEC%
}}\sim 6.962$ G. The condensate was loaded into a radially symmetric,
pancake-shaped trap with HO frequencies $(\nu _{x},\nu _{y},\nu
_{z})=(46,44,133)$ Hz, in the presence of the external magnetic field in the
$z$ direction, along which atomic magnetic dipoles are polarized, as shown
in Fig. \ref{fig-08}(A). Subsequently, the magnetic field was ramped down
to a value at which $a_{s}\approx a_{\mathrm{bg}}$, resulting in an angular
roton instability. Thus, the condensate evolved into a set of $N_{d}$
droplets, ranging between $2$ and $10$, which arranged themselves into a
triangular structure. As shown in Fig. \ref{fig-08}(B), $N_{d}$ shows a
linear dependence on the total number of atoms number, with $N/N_{d}\approx
1750$. The droplets strongly repel each other, maintaining distance $%
d\approx 3.0$ or $3.3$ $\mathrm{\mu }$m for $N_{d}=2$ and $N_{d}>2$,
respectively.

The spatial density distribution of the dysprosium condensate is
characterized by its Fourier transform, $S(k)$, which features a local
maximum at $k=2\pi /d\approx 2.5$ $\mathrm{\mu }$m, where $k=\sqrt{%
k_{x}^{2}+k_{y}^{2}}$. The spectral weight,
\begin{equation}
\mathrm{SW}=\sum_{k=1.5~\mathrm{\mu m}^{-1}}^{5~\mathrm{\mu m}^{-1}}S(k),
\label{SW}
\end{equation}%
accounts for the strength of the structured states. It is subject to
normalization $SW_{\mathrm{BEC}}=1$ for the entire condensate. To explore
properties of the spectral weight, a set of experimental data was collected,
as shown in Fig. \ref{fig-08}(D). The dysprosium condensate was generated
close to the FR at $a_{s}\approx a_{dd}$, and then ramped down to a target
value of the magnetic field, $6.860$ G, with a constant speed, see the red
arrow in Fig. \ref{fig-08}(C). This was followed by a waiting stage,
lasting for $20$ ms. Then, the magnetic field was increased back to the high
value at which the BEC was originally created, see the green arrow in Fig. %
\ref{fig-08}(C). In the course of the experiment, atomic samples were
imaged \emph{in situ}, and the corresponding spectral weights were
calculated as per Eq. (\ref{SW}), see Fig. \ref{fig-08}(D). A well-defined
hysteresis was thus observed, comparing the stages of the reduction of the
magnetic field and return back to the original value, indicating that the
system features bistability in the transition region.

To reveal the nature of the droplets observed in Ref. \cite{Kadau-Nat-2016},
they were trapped in a waveguide \cite{FB-PRL-2016}, which imposes a
prolate cigar-like shape with aspect ratio $\lambda \simeq 8$. The magnetic
field was ramped from $B_{\mathrm{BEC}}=6.962$ G to $B_{1}=6.656$ G during $%
1 $ ms. When quenching up $\varepsilon _{dd}$, the system, instead of
collapsing, forms a metastable state composed of droplets whose number is $%
1\leq N_{d}\leq 6$. For the case of $N_{d}\geq 2$, an average separation
between the droplets was measured to be $d=2.5\ \mathrm{\mu }$m. The
lifetime of these droplets is on order of hundreds of milliseconds, which is
much larger than that of the two-component QDs supported by contact
interactions. In addition, following the quench of the magnetic field, the
expanding droplets overlap, as their size become comparable to or larger
than the distance between them, which leads to the appearance of
interference fringes. The observation of the fringes
indicates that the individual droplets are phase-coherent objects, which
was also observed in one-dimensional droplet arrays \cite{Wenzel-2017-PRA}. In this connection, it is
relevant to mention that global phase coherence has been demonstrated in the supersolid state of matter,
which was recently realized in several experiments \cite{Tanzi-2019-PRL, FB-2019-PRX, Chomaz-2019-PRX, Guo-2019-Nat, Tanzi-2019-Nat, Natale-2019-PRL, Hertkorn-2019-PRL}.

\subsection{Quantum droplets in the condensate of erbium}

\begin{figure}[tbp]
\centering{\includegraphics[scale=0.7]{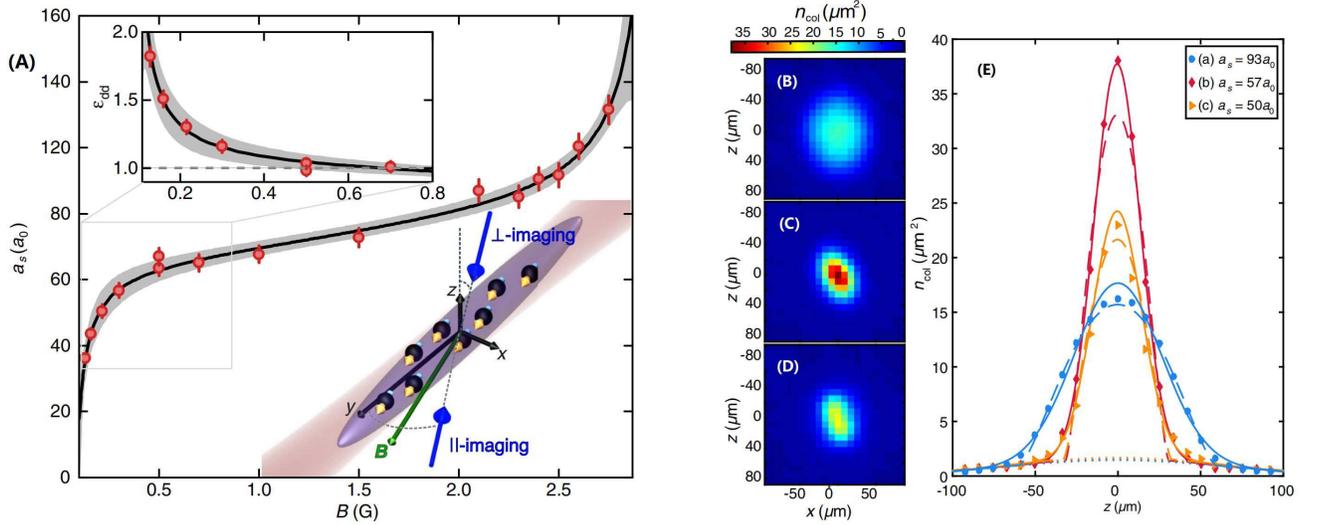}}
\caption{(A) Scattering length in $^{166}$Er versus magnetic field $B$. Data
points (red circles) are extracted from spectroscopic measurements, and the
solid line is a fit to the data set, with its statistical uncertainty (the
gray shaded region). The upper inset is the plot of $\protect\varepsilon %
_{dd}$ versus $B$. (C-E) Density profiles for different $a_{s}$ in the
BEC-QD crossover. (E) Lines show central cuts of the 2D bimodal fitting, the
solid (dashed) lines showing the two-Gaussian (MF-TF plus Gaussian)
distributions and the dotted lines represent the corresponding broad thermal
Gaussian part. The results are presented as per Ref. \protect\cite{Chomaz-PRX-2016}}
\label{fig-09}
\end{figure}
\begin{figure}[tbp]
\centering{\includegraphics[scale=0.73]{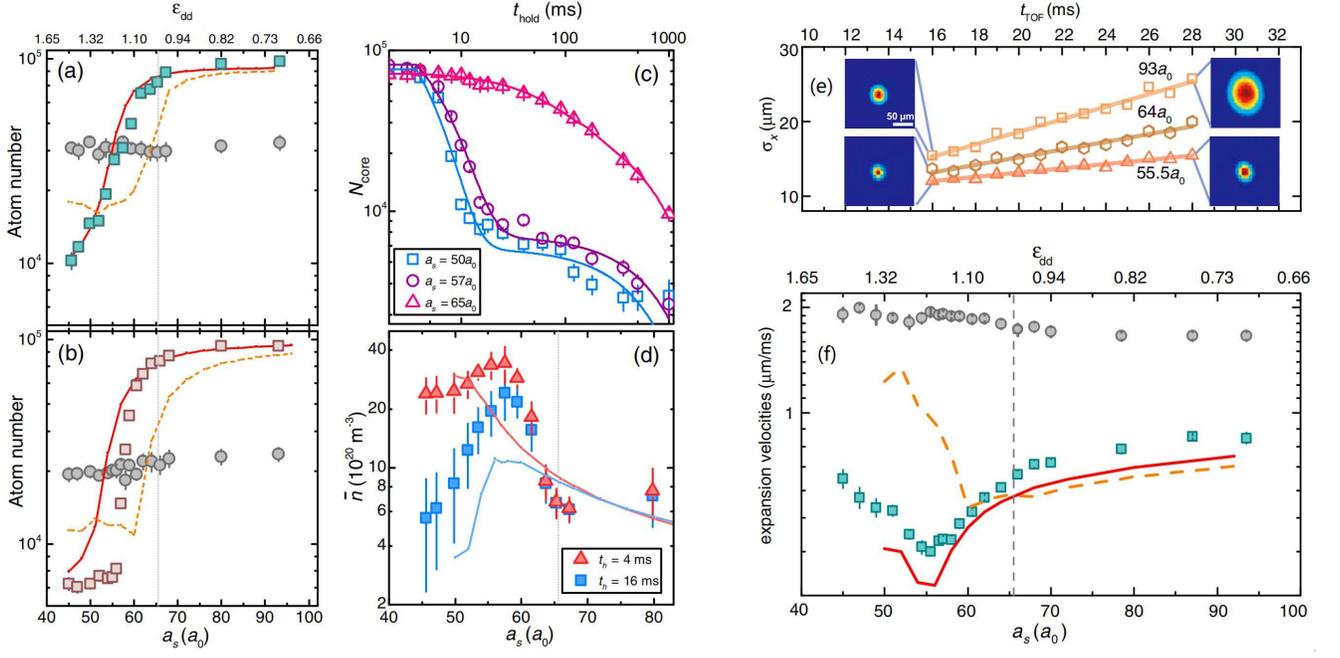}}
\caption{(a) and (b): Measured $N_{\mathrm{core}}$ (squares) and $N_{\mathrm{%
th}}$ (circles) versus $a_{s}$ after the action of (a) the nonadiabatic ($%
t_{r}=10$ ms, $t_{h}=8$ ms) and (b) adiabatic ($t_{r}=45$ ms, $t_{h}=0$ ms)
ramp. The data set shows better agreement with the theory including the LHY
term (the solid line), as compared to the MF theory (the dashed line). (c)
Time decay of $N_{\mathrm{core}}$ for $a_{s}=65a_{0}$ (triangles), $57a_{0}$
(circles), and $50a_{0}$ (squares) after quenching $a_{s}$ ($t_{r}=10$ ms).
(d) The mean \emph{in situ} density in the core, $\bar{n}$, for $t_{h}=4$
ms (triangles) and $16$ ms (squares), as a function of $a_{s}$. Solid lines
show results of the real-time simulation including the LHY correction, for $%
t_{h}=0$ ms (red) and $t_{h}=25$ ms (blue). (e) The TOF evolution of width $%
\protect\sigma _{x}$ of the high-density component for $a_{s}=93a_{0}$
(squares), $64a_{0}$ (circles), and $55.5a_{0}$ (triangles). (f) The
expansion velocity, $v_{x}$, as a function of $a_{s}$ (squares). For
comparison, the $a_{s}$-independent expansion velocities of the thermal
component are also shown (circles). The experimental data set is in very
good agreement with simulations of the parameter-free theory including the
LHY term (the solid line), while predictions of the MF-only theory (the
dotted line) are in discrepancy with the numerical findings. The results are
presented as per Ref. \protect\cite{Chomaz-PRX-2016}, where the experiments
were run with BEC of $^{166}$Er atoms.}
\label{fig-10}
\end{figure}

Erbium is another atomic species with permanent magnetic moment, which is
appropriate for the realization of dipolar BEC. Isotope $^{166}$Er was
employed in Ref. \cite{Chomaz-PRX-2016} to investigate the BEC-QD crossover.
The value of the respective background scattering length $a_{\mathrm{bg}}$
is comparable to the effective dipolar length, $a_{dd}=65.5a_{0}$, which
makes it easy to realize condition $\varepsilon _{dd}\simeq 1$, see Eq. (\ref%
{dd}). These atoms also feature a convenient set of FRs at ultra-low
magnetic field values. The dependence of the scattering length $a_{s}$ on $B$
is demonstrated in Fig. \ref{fig-09}(A). As shown in the upper inset, one
can easily enter the range of $\varepsilon _{dd}>1$, in which QDs may be
observed.

The condensate was prepared at $B=1.9$ G, corresponding to $a_{s}=81a_{0}$.
Following the evaporative cooling procedure, the magnetic field was reduced
to $0.8$ G [corresponding to $a_{s}=67a_{0}$, via the FR], with the atomic
magnetic moments polarized along the weak-trapping axis. Finally, $B$ was
ramped down to a target value in the course of time $t_{r}$, which is
followed by wait time $t_{h}$. Then, an absorption image of the gas was
taken, after time-of-flight (TOF) $t_{\mathrm{TOF}}$. Figures \ref{fig-09}%
(B-D) display typical absorption images of the density profiles for $%
t_{r}=10 $ ms (quenching), $t_{h}=6$ ms, $t_{\mathrm{TOF}}=27$ ms, and
different values of $a_{s}$. In particular, in the case of $\varepsilon
_{dd}>1$, as shown in Fig. \ref{fig-09}(E), the density distribution is
close to that predicted by the Thomas-Fermi (TF) approximation, which
neglects the kinetic-energy term in GPE. The distribution of thermal atoms,
see dotted lines in Fig. \ref{fig-09}(E), is different from that in the
central core, and remains mainly unaffected by the change of $a_{s}$.
Collective oscillations of the coherent gas cloud is intimately related to
the origin of the stabilization mechanism. In this work, the axial mode,
which is the lowest-lying excitation in the system above the dipolar mode,
was experimentally studied for both adiabatic and nonadiabatic ramps of the
magnetic field. For both ramps, the results highlight a qualitative
agreement with the theoretical predictions including the LHY term, revealing
the fact that the LHY correction plays an essential role in stabilizing the
system.

As said above, quantum fluctuations are expected to stabilize the system and
help forming the droplets. However, 3B losses favor lower densities. The
interplay between quantum fluctuations and 3B losses in the BEC-to-QD
crossover is of great interest. Numbers of atoms in both the central-core ($%
N_{\mathrm{core}}$) and thermal ($N_{\mathrm{th}}$) components are shown, as
a function of $a_{s}$, in Figs. \ref{fig-10}(a) and (b), respectively,
following the action of the nonadiabatic and adiabatic ramps. Both cases
show a similar evolution. When the magnetic field is ramped down, the~number
of atoms in the central core remains constant for $\varepsilon _{dd}>1$,
then drops dramatically around $\varepsilon _{dd}\simeq 1$, and finally
curves up at lower $a_{s}$. In contrast to that, $N_{\mathrm{th}}$ shows
weak dependence on $a_{s}$, confirming a picture in which dynamics of the
thermal and condensed components are uncoupled. The observed evolution of $%
N_{\mathrm{core}}$ matches well with the theoretical calculation including
the LHY correction (solid lines), but deviates from the one performed in the
absence of the LHY term (dashed lines). The time evolution of $N_{\mathrm{%
core}}$ for various $a_{s}$ in the droplet regime is displayed in Fig. \ref%
{fig-10}(c), where $N_{\mathrm{core}}$ shows fast decay in the interval of $%
3.5<t<25$ ms, indicating that atoms are ejected from the high-density core
through 3B losses. The steepness of this fast decay critically depends on $%
a_{s}$. The mean \emph{in situ} density $\bar{n}$ of the high-density
component in the BEC-QD crossover is extracted with the help of the general
3B-loss relation, see further details in Ref. \cite{Chomaz-PRX-2016}. As
shown in Fig. \ref{fig-10}(d), one can see that the mean density $\bar{n}$
attains a maximum at the threshold, with $a_{s}\simeq a_{dd}$, showing a
quantitative agreement with numerical simulations including the LHY
correction.

As a self-trapped state, the QD is expected to demonstrate its
characteristic in the expansion regime. Typical examples of the TOF
evolution of width $\sigma _{x}$ of the high-density core are shown in Fig. %
\ref{fig-10}(e). It is observed that, in the case of $\varepsilon _{dd}>1$,
the atomic cloud exhibits clear slowing-down of the expansion dynamics. The
expansion velocity $v_{x}$ is extracted by fitting the data to $\sigma
_{x}(t_{\mathrm{TOF}})=\sqrt{\sigma _{x,0}^{2}+v_{x}^{2}t_{\mathrm{TOF}}^{2}}
$. The dependence of velocity $v_{x}$ of the expansion of the high-density
core on $a_{s}$ is demonstrated in Fig. \ref{fig-10}(f). In the droplet
regime $v_{x}$ gets a minimum at about $56a_{0}$ ($\varepsilon _{dd}\sim 1.17
$), and grows when $a_{s}$ gets far away from this minimum point. This
behavior cannot be explained by the MF theory. On the other hand,
simulations with the LHY correction reproduce the results produced by
experimental measurements, see the solid line in Fig. \ref{fig-10} (f).

\section{Theoretical results: stable quantum droplets with embedded vorticity}

\subsection{Three-dimensional vortex rings}

\begin{figure}[tbp]
\centering{\includegraphics[scale=1]{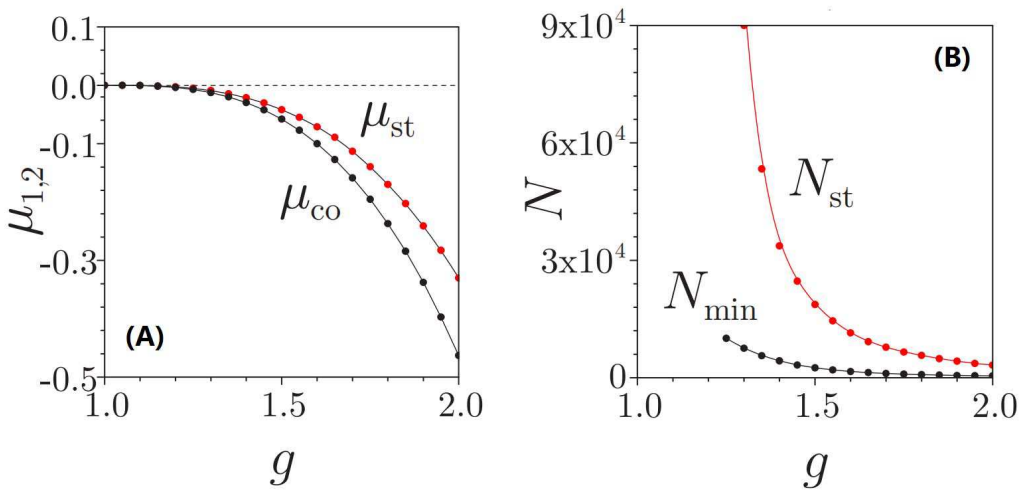}}
\caption{(A): In the framework of Eqs. (\protect\ref{3D}), 3D symmetric
vortex rings, in the form Eq. (\protect\ref{m12}), with $m_{1}=m_{2}=1$ and $%
g_{\mathrm{LHY}}=0.5$, are stable in region $\protect\mu _{\mathrm{co}}<%
\protect\mu <\protect\mu _{\mathrm{st}}$ in the plane of the relative
strength of the intercomponent attraction, $g$, and equal chemical
potentials $\protect\mu _{1}=\protect\mu _{2}$. Panel (B) shows the minimum
norm, $N_{\min }$, above which the vortex rings exist, and the boundary
value, $N_{\mathrm{st}}$, above which they are stable in the plane of $%
\left( g,N\right) $. The results are displayed as per Ref. \protect\cite%
{Kartashov-PRA-2018}}
\label{fig-11}
\end{figure}

Quantum droplets observed in experiments outlined above are fundamental
modes, which do not carry any vorticity. It is natural to expect that vortex
(alias spinning) modes may offer an opportunity to study more sophisticated
properties of the QD state of matter \cite{Malomed-2019-PD}. Thus far, all
studies of QDs with embedded vorticity were performed solely in the
theoretical form. In particular, in Ref. \cite{Cidrim-PRA-2018} it was
demonstrated that QD solutions with embedded vorticity exist in the model of
the single-component dipolar condensate, but they all are unstable, hence
physically irrelevant.

On the other hand, it was found that models of binary condensates with
contact interactions, based on systems of LHY-amended GPEs readily give rise
to \emph{stable} vortex states. In particular, Ref. \cite{Kartashov-PRA-2018}
addressed the 3D system for the two-component wave function $\psi _{1,2}$.
In the scaled form, the system takes the form of
\begin{eqnarray}
&&i\frac{\partial \psi _{1}}{\partial t}=-\frac{1}{2}\nabla ^{2}\psi
_{1}+(\left\vert \psi _{1}\right\vert ^{2}+g_{\mathrm{LHY}}\left\vert \psi
_{1}\right\vert ^{3})\psi _{1}-g\left\vert \psi _{2}\right\vert ^{2}\psi
_{1},  \notag \\
&&i\frac{\partial \psi _{2}}{\partial t}=-\frac{1}{2}\nabla ^{2}\psi
_{2}+(\left\vert \psi _{2}\right\vert ^{2}+g_{\mathrm{LHY}}\left\vert \psi
_{2}\right\vert ^{3})\psi _{2}-g\left\vert \psi _{1}\right\vert ^{2}\psi
_{2},  \label{3D}
\end{eqnarray}%
where the strength of the cubic self-repulsion in each component is scaled
to be $1$, while $g>0$ is the relative strength of the inter-component
attraction. The LHY repulsion is characterized by coefficient $g_{\mathrm{LHY%
}}\simeq (128/3)\sqrt{2/\pi a_{s}^{3/2}}$, where $a_{s}$ is the
intra-component scattering length. The corresponding stationary solutions
for vortex droplets with chemical potentials $\mu _{1,2}$ and integer
topological charges $m_{1,2}$ of the components are looked for, in
cylindrical coordinates $(\rho ,\theta ,z)$, as
\begin{equation}
\psi _{1,2}=u_{1,2}(\rho ,z)\exp (im_{1,2}\theta -i\mu _{1,2}t).  \label{m12}
\end{equation}%
with real stationary wave functions $u_{1,2}$ obeying equations
\begin{equation}
\mu u_{1,2}+\frac{1}{2}\left( \frac{\partial ^{2}}{\partial \rho ^{2}}+\frac{%
1}{\rho }\frac{\partial }{\partial \rho }+\frac{\partial ^{2}}{z^{2}}-\frac{%
m_{1,2}^{2}}{\rho ^{2}}\right) u_{1,2}-\left( u_{1,2}^{2}+g_{\mathrm{LHY}%
}u_{1,2}^{3}\right) u_{1,2}+gu_{2,1}^{2}u_{1,2}=0.
\end{equation}

Stability regions for the 3D QDs with embedded vorticity, i.e., \textit{%
vortex rings}, which are symmetric with respect to the two components, with $%
\mu _{1}=\mu _{2}$ and $m_{1}=m_{2}=1$, are shown in Fig. \ref{fig-11}, both
in the $(\mu ,g)$ and $(N,g)$ planes. As demonstrated in Fig. \ref{fig-11}%
(A), the vortex rings are stable in region $\mu _{\mathrm{co}}<\mu <\mu _{%
\mathrm{st}}$, where $\mu _{\mathrm{co}}$ is the cutoff value of the
chemical potential, below which no droplet can be found. Actually, $\mu _{%
\mathrm{co}}$ corresponds to the indefinitely broad QDs with a flat-top
shape and diverging integral norm (number of atoms), the existence of $\mu _{%
\mathrm{co}}$ being a general property of self-trapped states in models with
competing nonlinearities, such as the well-studied cubic-quintic combination
\cite{Bulgaria,Bulgaria2}. The stability interval of $\mu $ expands as $g$
increases, due to the fact that the value of $\mu _{\mathrm{co}}$ decreases
faster than the stability boundary $\mu _{\mathrm{st}}$. The stability
domain in the $(N,g)$ plane is shown in Fig. \ref{fig-11}(B). Value $N_{\min
}$ indicates the minimum number of atoms necessary for the formation of a
spinning droplet. The 3D vortex droplets are stable at $N>N_{\mathrm{st}}$,
while In the region of $N_{\min }<N<N_{\mathrm{st}}$ they exist but are
unstable. The stability-boundary value $N_{\mathrm{st}}$ rapidly increases
with the decrease of $g$, and diverges at $g=g_{\min }\approx 1.3$, below
which system (\ref{3D}) does not maintain stable spinning states.

For the symmetric state with double vorticity, of $m_{1}=m_{2}=2$, a narrow
stability region was obtained (not shown here in detail). It may happen that
stable higher-order vortices with $m_{1,2}\equiv m\geq 3$ exist too in this
3D system, but they were not found. This fact may be explained by scaling
\begin{equation}
N_{\mathrm{th}}^{\mathrm{(3D)}}\sim m^{6}  \label{min3D}
\end{equation}%
\ for the minimum number of atoms necessary for the existence of stable
vortex modes in this 3D model \cite{Li-PRA-2018}, as this very steep scaling
makes it very difficult to create such stable modes with $m\geq 3$. On the
other hand, QDs with \textit{hidden vorticity}, i.e., as defined in Ref.
\cite{Gammal}, antisymmetric states with opposite vorticities in the two
components,
\begin{equation}
m_{1}=-m_{2}=-1,  \label{hidden}
\end{equation}%
are completely unstable in the framework of Eqs. (\ref{3D}).

\subsection{Two-dimensional vortex rings and necklaces}

\subsubsection{Basic results}

\begin{figure}[tbp]
\centering{\includegraphics[scale=0.8]{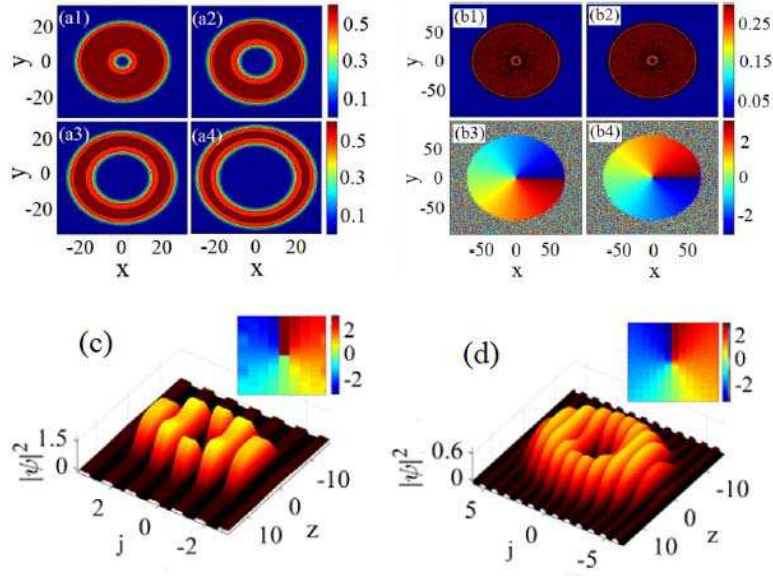}}
\caption{Panels (a1)-(a4) display density patterns of vortex QDs with $%
S=1,2,3,4$ and norm $N=1000$. Panels (b1-b4): Density and phase plots
showing opposite vorticities of the two components of a stable
hidden-vorticity mode, with $(N,4\protect\pi /g)=(8000,0.11)$, as a result
of the evolution simulated up to $t=10000$. (c,d) Examples of stable
on-site- and intersite-centered vortices, for $(g,C)=(0.48;0.1)$ and $%
(0.77,0.15)$. Panels (a1-a4) and (b1-b4) are borrowed from Ref. \protect\cite%
{Li-PRA-2018}, and panels (c,d) are borrowed from Ref. \protect\cite%
{Zhang-2019-PRL}. }
\label{vortexQD}
\end{figure}

In the framework of the reduction of the LHY-amended GPE system to the 2D
form, see Eq. (\ref{log}), families of stable QDs with embedded vorticity,
both explicit (identical in both components) and \textit{hidden}, defined as
per Eq. (\ref{hidden}), were explored in Ref. \cite{Li-PRA-2018}. The
extension of Eq. (\ref{log}) for two components, $\psi _{\pm }$, is
\begin{equation}
i\partial _{t}\psi _{\pm }=-\frac{1}{2}\nabla ^{2}\psi _{\pm }+\frac{4\pi }{g%
}\left( |\psi _{\pm }|^{2}-|\psi _{\pm }|^{2}\right) \psi _{\pm }+(|\psi
_{\pm }|^{2}+|\psi _{\mp }|^{2})\psi _{\pm }\ln (|\psi _{\pm }|^{2}+|\psi
_{\mp }|^{2}),  \label{loglog}
\end{equation}%
where $g>0$ is the coupling constant. It was found that Eq. (\ref{loglog})
gives rise to stable 2D vortex-ring solutions with explicit vorticity up to $%
m_{1}=m_{2}\equiv m=5$. An essential finding is that such QDs with embedded
vorticity are stable above a certain threshold value, $N_{\mathrm{th}}^{%
\mathrm{(2D)}}$, of the number of atoms, which scales as
\begin{equation}
N_{\mathrm{\min }}^{\mathrm{(2D)}}\sim m^{4}  \label{min2D}
\end{equation}%
with the increase of $m$, cf. the steeper scaling in the 3D model (\ref%
{min3D}). The possibility to find stable 2D vortices with $m\geq 3$ is a
consequence of the relatively mild scaling in Eq. (\ref{min2D}), in
comparison with Eq. (\ref{min3D}).

Vortex QDs in a similar 2D model, with GPEs including both the LHY
correction and spin-orbit coupling between the two components, were
considered too \cite{Li-2017-NJP}. In the latter case, QDs are states of the
\textit{mixed-mode} type, in terms of Ref. \cite{Ben Li}, with each
component including terms with vorticities $m=0$ and $+1$ or $-1$.

\subsubsection{Semidiscrete vortices}

Further, a \textit{semidiscrete} 2D system, which is constructed as an array
of quasi-1D cigar-shaped waveguides for QDs, was introduced in Ref. \cite%
{Zhang-2019-PRL}. In the 1D limit, the LHY correction to the 1D GPE\ is a
quadratic term with the attraction sign \cite{Petrov-PRL-2016}, on the
contrary to the repulsive quartic LHY term in Eq. (\ref{Petrov}) and the
alternating attraction-repulsion sign of the logarithmic factor in the 2D
equation (\ref{log}). The scaled form of the GPE system for this system
includes the MF cubic self-repulsion terms, competing with the LHY-induced
quadratic self-attraction:
\begin{equation}
i\partial \psi _{j}=-\frac{1}{2}\partial _{zz}\psi _{j}-\frac{C}{2}\left(
\psi _{j-1}-2\psi _{j}+\psi _{j}\right) +g|\psi _{j}|^{2}\psi _{j}-|\psi
_{j}|\psi _{j},  \label{semidiscrete}
\end{equation}%
where $C>0$ is the hopping rate which provides linear coupling between
adjacent between cores (waveguides), the strength of the quadratic LHY term
is normalized to be one, and $g>0$ is the strength of the cubic
self-repulsion. This system gives rise to stable vortical \textit{%
semidiscrete} QDs, with the winding number (embedded vorticity) up to $m=5$.
Among them, there are two different species of stable semidiscrete vortex QD
with $m=1$, of the on-site-centered and inter-site-centered types, with the
vorticity pivot located, respectively, at a lattice site or between two
sites, see examples in Figs. \ref{vortexQD}(c,d)]. Stable
inter-site-centered vortices are found only when the hopping rate between
adjacent cores, $C$, is very small, and they do not exist with $m\geq 2$. On
the other hand, stable on-site-centered vortices were found for arbitrary
values of $C$, with embedded vorticities $1\leq m\leq 5$, i.e., in the same
range as indicated above for the 2D continuum model based on Eq. (\ref%
{loglog}).

\subsubsection{Necklace clusters}

Necklace patterns, built as ring-shaped clusters of solitons, usually appear
to be unstable patterns, because interactions between adjacent soliton split
the necklaces into sets of separating solitons. Nevertheless, Ref. \cite%
{Kartashov-2019-PRL} has demonstrated a possibility to construct robust
necklace clusters, composed of fundamental (zero-vorticity) 2D droplets, in
the framework of the model based on Eq. (\ref{loglog}). The cluster is built
of $n$ identical QDs placed on a ring of radius $R$, with phase difference $%
2\pi m/n$ between adjacent QDs, where $m$ is the overall vorticity imprinted
onto the cluster. The evolution of the cluster is driven by the initial
radius, $R$, and vorticity $m$, ranging from contraction to rotation or
expansion. As a result, the necklaces which realize an energy minimum
feature remarkable robustness, while they are strongly unstable in the
framework of the 2D GPE with usual cubic nonlinearity.

Unlike the imprinted vorticity considered above, an alternative way to
construct vortex clusters carrying angular momentum is offered by the ground
state of a rotating trapped binary BEC, as recently demonstrated in Ref.
\cite{Tengstrand-2019}. Such\ a system, with attractive inter-species and
repulsive intra-species interactions, is confined in a shallow HO harmonic
trap with an additional repulsive Gaussian potential placed at the center.
If the LHY correction is taken into account, it helps to stabilize the
patterns. Numerical simulations have produced rotating necklace-like
patterns composed of a few local vortices with topological charge $m=1$.
Traces of these patterns persist in the expanding condensate if it is
released into free space when the weakly confining HO trapping potential is
switched off.

\section{Two-dimensional vortex modes trapped in a singular potential}

\subsection{Formulation of the problem}

The quantum collapse is a well-known peculiarity in quantum mechanics:
nonexistence of the GS in 3D and 2D linear Schr\"{o}dinger equations with
attractive potential
\begin{equation}
U(r)=-\left( U_{0}/2\right) r^{-2},  \label{U}
\end{equation}%
where $U_{0}>0$ is the strength of the pull to the center \cite{LL}. In 3D,
the collapse occurs when $U_{0}$ exceeds a finite critical value, while in
2D the collapse happens at any $U_{0}>0$. In both 3D and 2D cases, the
potential represents attraction of a particle, carrying a permanent electric
dipole moment, to a central charge \cite{HS1}. In 2D, the same potential (%
\ref{U}) may be realized as attraction of a magnetically polarizable atom to
an electric current (e.g., an electron beam) directed transversely to the
system's plane, or the attraction of an electrically polarizable atom to a
uniformly charged transverse thread.

A fundamental issue is regularization of the setting, aiming to create a
missing GS. A solution was proposed in Ref. \cite{HS1}, replacing the 3D
linear Schr\"{o}dinger equation by the GPE for a gas of dipole particles
pulled to the center by potential (\ref{U}) and stabilized by repulsive
contact interactions, while the long-range DDI\ between the particles amount
to a renormalization of the contact interaction \cite{HS1}. It was thus
found that, in the framework of the MF\ approximation, the 3D GPE creates
the missing GS for arbitrarily large $U_{0}$. Further, it was demonstrated
that, in terms of the many-body quantum theory, the GS, strictly speaking,
does not exist in the same setting, but the interplay of the pull to the
center and contact repulsion gives rise to a metastable state, separated
from the collapsing state by a tall potential barrier \cite{Gregory}.

The situation is more problematic in 2D, as the usual cubic nonlinearity,
which represents the contact repulsion in the MF approximation, is not
strong enough to create the GS. The problem is that the MF wave function, $%
\psi (r)$, produced by GPE, gives rise to the density, $\left\vert \psi
(r)\right\vert ^{2}$, diverging $\sim r^{-2}$ at $r\rightarrow 0$, in 3D and
2D alike. In terms of the integral norm,%
\begin{equation}
N=\lim_{r_{\mathrm{cutoff}}\rightarrow 0}\left[ \left( 2\pi \right)
^{D-1}\int_{r_{\mathrm{cutoff}}}^{\infty }\left\vert \psi (r)\right\vert
^{2}r^{D-1}dr\right] ,  \label{N}
\end{equation}%
where $D=3$ or $2$ is the dimension, the density singularity $\sim r^{-2}$
is integrable in 3D, while it gives rise to a logarithmic divergence in 2D,
\begin{equation}
N\sim \ln \left( r_{\mathrm{cutoff}}^{-1}\right) .  \label{cutoff}
\end{equation}%
The analysis of GPE demonstrates that a self-repulsive nonlinear term
stronger than cubic, i.e., $|\Psi |^{\alpha -1}\Psi $ with $\alpha >3$,
gives rise to the density with singularity $|\psi (r)|^{2}\sim r^{-4/(\alpha
-1)}$. Therefore, any value $\alpha >3$ provides convergence of the 2D
integral norm.

Thus, a solution for the regularization of the 2D setting may be offered by
the quintic defocusing nonlinearity \cite{HS1}, with $\alpha =5$. It
accounts for three-body repulsive interactions in the bosonic gas \cite%
{Abdullaev}, although the realization of this feature is the fact that
three-particle collisions give rise to losses, kicking out particles from
the condensate \cite{loss1}. On the other hand, the LHY-induced quartic
self-repulsive term in Eq. (\ref{Petrov}) may also be used for the
stabilization of the 2D setting under the action of potential (\ref{U}) \cite%
{Shamriz}, provided that the confinement in the transverse direction is
realistic (not extremely tight). As mentioned above, in this case the LHY
effect is accounted for by the quartic term, added to the effectively
two-dimensional GPE. In this connection, it is relevant to note that, if the
\textquotedblleft fully 2D" equation (\ref{log}), corresponding to the
ultra-tight confinement, is insufficient to create a GS with a convergent
norm in 2D. Indeed, in this case the analysis yields a density singularity $%
\left\vert \psi \right\vert ^{2}\sim r^{-2}/\ln \left( r^{-1}\right) $ at $%
r\rightarrow 0$, hence the 2D integral (\ref{N}) is still diverging,
although extremely slowly, $N\sim \ln \left( \ln \left( r_{\mathrm{cutoff}%
}^{-1}\right) \right) $, cf. Eq. (\ref{cutoff}).

Thus, the relevant two-dimensional LHY-amended GPE equation, including
potential (\ref{U}), takes the following form in the scaled notation \cite%
{Shamriz}:
\begin{equation}
i\frac{\partial \psi }{\partial t}=-\frac{1}{2}\left( \frac{\partial
^{2}\psi }{\partial r^{2}}+\frac{1}{r}\frac{\partial \psi }{\partial r}+%
\frac{1}{r^{2}}\frac{\partial ^{2}\psi }{\partial \theta ^{2}}\right) -\frac{%
U_{0}}{2r^{2}}\psi +\sigma \left\vert \psi \right\vert ^{2}\psi +|\psi
|^{3}\psi ,  \label{psi2d}
\end{equation}%
which is written in polar coordinates $\left( r,\theta \right) $,
coefficient $\sigma $ accounting for the residual MF nonlinearity. In fact,
the rescaling makes it possible to set $\sigma =\pm 1$ or $0$. The case of $%
\sigma =0$ corresponds to the (nearly) exact cancellation between the
intra-component repulsion and inter-component attraction, while all the
nonlinearity is represented by the LHY-induced quartic term, cf. Ref. \cite%
{only}. While Ref. \cite{Shamriz} addressed Eq. (\ref{psi2d}) in its general
form, we here focus on the most fundamental case of $\sigma =0$.

\subsection{Analytical considerations}

Stationary solutions to Eq. (\ref{psi2d}) with chemical potential $\mu <0$
and integer vorticity $m$ are looked for as%
\begin{equation}
\psi \left( r,t\right) =\exp \left( -i\mu t+im\theta \right) u(r),
\label{psichi2D}
\end{equation}%
with real radial function satisfying the equation%
\begin{equation}
\mu u=-\frac{1}{2}\left( \frac{d^{2}u}{dr^{2}}+\frac{1}{r}\frac{du}{dr}+%
\frac{U_{m}}{r^{2}}u\right) +u^{4},  \label{chi2D}
\end{equation}%
where, as said above, we set $\sigma =0$, and define a renormalized
potential strength,%
\begin{equation}
U_{m}\equiv U_{0}-m^{2}.  \label{Ul}
\end{equation}%
Simple corollaries of Eq. (\ref{chi2D}) are scaling relations which show the
dependence of the solution on $\mu $:%
\begin{equation}
u\left( r;\mu \right) =\left( -\mu \right) ^{1/3}u\left( \left( -\mu \right)
^{-1/2}r;\mu =-1\right) ,  \label{scaling}
\end{equation}%
\begin{equation}
N(\mu )=\left( -\mu \right) ^{-1/3}N(\mu =-1).  \label{scalingN}
\end{equation}

A convenient substitution,
\begin{equation}
\psi \left( r,\theta ,t\right) \equiv r^{-2/3}\varphi \left( r,\theta
,t\right) ,u(r)\equiv r^{-2/3}\chi (r),  \label{uchi}
\end{equation}%
transforms Eqs. (\ref{psi2d}) and (\ref{chi2D}) into
\begin{eqnarray}
i\frac{\partial \varphi }{\partial t} &=&-\frac{1}{2}\left[ \frac{\partial
^{2}}{\partial r^{2}}-\frac{1}{3r}\frac{\partial }{\partial r}+\frac{\left(
U_{0}+4/9\right) }{r^{2}}+\frac{1}{r^{2}}\frac{\partial ^{2}}{\partial
\theta ^{2}}\right] \varphi  \notag \\
&&+\sigma \frac{|\varphi |^{2}\varphi }{r^{4/3}}+\frac{|\varphi |^{3}\varphi
}{r^{2}},  \label{varphi}
\end{eqnarray}%
\begin{equation}
\mu \chi =-\frac{1}{2}\left[ \frac{d^{2}\chi }{dr^{2}}-\frac{1}{3r}\frac{%
d\chi }{dr}+\frac{\left( U_{m}+4/9\right) }{r^{2}}\chi \right] +\sigma \frac{%
\chi ^{3}}{r^{4/3}}+\frac{\chi ^{4}}{r^{2}}.  \label{chi}
\end{equation}%
The expansion of the solution to Eq. (\ref{chi}) at $r\rightarrow 0$ yields
\begin{equation}
\chi (r)=\left[ \frac{1}{2}\left( U_{m}+\frac{4}{9}\right) \right]
^{1/3}\left( 1+\frac{2\mu }{3U_{m}}r^{2}\right) +O\left( r^{4}\right) ,
\label{expansion}
\end{equation}
which is valid for $U_{m}>0$. In the interval of%
\begin{equation}
-4/9<U_{m}<0  \label{-4/9}
\end{equation}%
(the meaning of this interval is explained below), the quadratic term in Eq.
(\ref{expansion}) is replaced, as the leading correction, by
\begin{equation}
\mathrm{const}\cdot r^{\beta },\beta =\frac{2}{3}+\sqrt{\frac{16}{9}+3U_{m}}%
<2,  \label{beta}
\end{equation}
where $\mathrm{const}$ remains indefinite, in terms of the expansion at $%
r\rightarrow 0$. Exactly at $U_{m}=0$, Eq. (\ref{expansion}) is replaced by
\begin{equation}
\chi (r;U_{m}=0)=\left( \frac{2}{9}\right) ^{1/3}\left( 1+\frac{3\mu }{4}%
r^{2}\ln \left( \frac{r_{0}}{r}\right) \right) +~...,  \label{expansionU=0}
\end{equation}%
where constant $r_{0}$ is also indefinite. In all the cases, it follows from
Eq. (\ref{uchi}) that the singular form of the density at $r\rightarrow 0$
is
\begin{equation}
u^{2}(r)\approx \left[ \frac{1}{2}\left( U_{m}+\frac{4}{9}\right) \right]
^{2/3}r^{-4/3},  \label{singular}
\end{equation}%
with which the 2D norm (\ref{N}) \emph{converges}.

The asymptotic form of the solution, given by Eq. (\ref{expansion}), is
meaningful if it yields $\chi (r)>0$ [otherwise, the derivation of Eq. (\ref%
{chi2D}) from Eq. (\ref{psi2d}) is irrelevant], i.e., for $U_{m}>0$, as well
as for \emph{weakly negative} values of the effective strength of the
central potential belonging to interval (\ref{-4/9}). In the limit of $%
U_{0}+4/9\rightarrow +0$, Eq. (\ref{chi}) gives rise to an \emph{%
asymptotically exact} solution:%
\begin{equation}
\chi (r;U_{m}+4/9\rightarrow 0)=\frac{\sqrt{3}\Gamma (1/3)}{\pi }\left[ -%
\frac{\mu }{4}\left( U_{m}+\frac{4}{9}\right) \right] ^{1/3}r^{2/3}K_{2/3}%
\left( \sqrt{-2\mu }r\right) ,  \label{exact}
\end{equation}%
where $\Gamma (1/3)\approx \allowbreak 2.68$ is the value of the
Gamma-function, and $K_{2/3}$ is the modified Bessel function of the second
kind. The substitution of this expression in Eqs. (\ref{uchi}) and (\ref{N})
yields the respective value of the norm,
\begin{equation}
N\left( U_{m}+4/9\rightarrow 0\right) =\frac{\Gamma ^{2}(1/3)}{\sqrt{3}}%
\frac{\left( U_{m}+4/9\right) ^{2/3}}{\left( -2\mu \right) ^{1/3}},
\label{exactN}
\end{equation}%
which agrees with scaling relation (\ref{scalingN}).

The counter-intuitive finding that the bound state may exist under the
combined action of the defocusing quartic nonlinearity and effectively
repulsive potential in interval (\ref{-4/9}), which was first reported in
Ref. \cite{HS1}, is explained in detail in Refs. \cite{Shamriz} and \cite%
{sing-sol}). This property is specific for singular bound states (which are
physically relevant ones, as they produce the convergent norm).

In the limit of $r\rightarrow \infty $, the asymptotic form of the solution
to Eq. (\ref{chi}) is%
\begin{equation}
\chi (r)\approx \chi _{0}r^{1/6}\exp \left( -\sqrt{-2\mu }r\right) ,
\label{chi0}
\end{equation}%
where $\chi _{0}$ is an arbitrary constant. A global picture of the
nonlinear modes is produced by the TF approximation, which neglects
derivatives in Eq. (\ref{chi}):%
\begin{equation}
\chi _{\mathrm{TF}}(r)=\left\{
\begin{array}{c}
\left[ \left( U_{m}+4/9\right) /2+\mu r^{2}\right] ^{1/3},~\mathrm{at}%
~~r<r_{0}\equiv \sqrt{-\left( U_{m}+4/9\right) /\left( 2\mu \right) }, \\
0,~\mathrm{at}~~r>r_{0}~.%
\end{array}%
\right.  \label{TF}
\end{equation}%
In the limit of $r\rightarrow 0$, Eq. (\ref{TF}) yields the same exact value
of $\chi (r=0)=\left[ \left( U_{m}+4/9\right) /2\right] ^{1/3}$ as given by
Eq. (\ref{expansion}). On the other hand, the TF\ approximation predicts a
finite radius $r_{0}$ of the mode, neglecting the exponentially decaying
tail at $r\rightarrow \infty $, cf. Eq. (\ref{chi0}).

The TF approximation makes it possible to calculate the corresponding $N(\mu
)$ dependence:%
\begin{equation}
N_{\mathrm{TF}}(\mu )=2\pi \int_{0}^{r_{0}}\left[ r^{-2/3}\chi _{\mathrm{TF}%
}(r)\right] ^{2}rdr=C\frac{U_{m}+4/9}{(-\mu )^{1/3}},  \label{NTF}
\end{equation}%
with $C\equiv \pi \int_{0}^{1}\left( x^{-2}-1\right) ^{2/3}xdx\approx 3.80$,
which complies with the exact scaling relation (\ref{scalingN}). The TF
approximation is quite accurate for sufficiently large values of $U_{m}$.
For instance, at $U_{m}=10$ and $\mu =-1$, a numerically found value of the
norm is $N_{\mathrm{num}}\approx 41.05$, while its TF-predicted counterpart
is $N_{\mathrm{TF}}\approx 39.68$.

\subsection{Vortices}

Usually, the presence of integer vorticity $m\geq 1$ implies that the
amplitude vanishes at $r\rightarrow 0$ as $r^{m}$, which is necessary
because the phase of the vortex field is not defined at $r=0$. However, the
indefiniteness of the phase is also compatible with the amplitude \emph{%
diverging} at $r\rightarrow 0$. In the linear equation, this divergence has
the asymptotic form of the standard singular Bessel's (alias Neumann's)
cylindrical function, $Y_{l}(r)\sim r^{-m}$, which makes the respective 2D
state unnormalizable for all $m\geq 1$. However, in the present system,
similar to Ref. \cite{HS1}, Eqs. (\ref{uchi}), (\ref{expansion}), (\ref{beta}%
) and (\ref{expansionU=0}) demonstrate that the interplay of the central
potential and quartic nonlinearity reduces the divergence of the amplitude
function to the level of $r^{-2/3}$, for any $m$, thus maintaining the
normalizability of the states under the consideration.

Stationary solutions of Eq. (\ref{chi}) are not essentially different for $%
m=0$ (the GS) and $m\geq 1$ (vortices). A real difference is revealed by the
analysis of their stability. Computation of stability eigenvalues for modes
of small perturbations and direct simulations of the perturbed evolution
demonstrate that all GSs, including those in the ``counter-intuitive"
interval (\ref{-4/9}, are \emph{completely stable} \cite{Shamriz}.

The situation is different for the vortices. The analysis of the linearized
equations for small perturbations leads to an exact result: they are stable
at%
\begin{equation}
U_{0}\geq \left( U_{0}\right) _{\mathrm{crit}}^{(m)}=(7/9)\left(
3m^{2}-1\right) ,  \label{lstability}
\end{equation}%
while at $U_{0}<$ $\left( U_{0}\right) _{\mathrm{crit}}^{(m)}$ the vortices
are unstable against a perturbation eigenmode which drives the vortex' pivot
out of the central position. This prediction, including the particular
values of $\left( U_{0}\right) _{\mathrm{crit}}^{(m)}$, was accurately
corroborated by numerical computation of stability eigenvalues, as well as
by direct simulations of perturbed evolution for $m=1$ and $2$ \cite{Shamriz}%
.

\begin{figure}[tbp]
\subfigure[]{\includegraphics[width=3.2in]{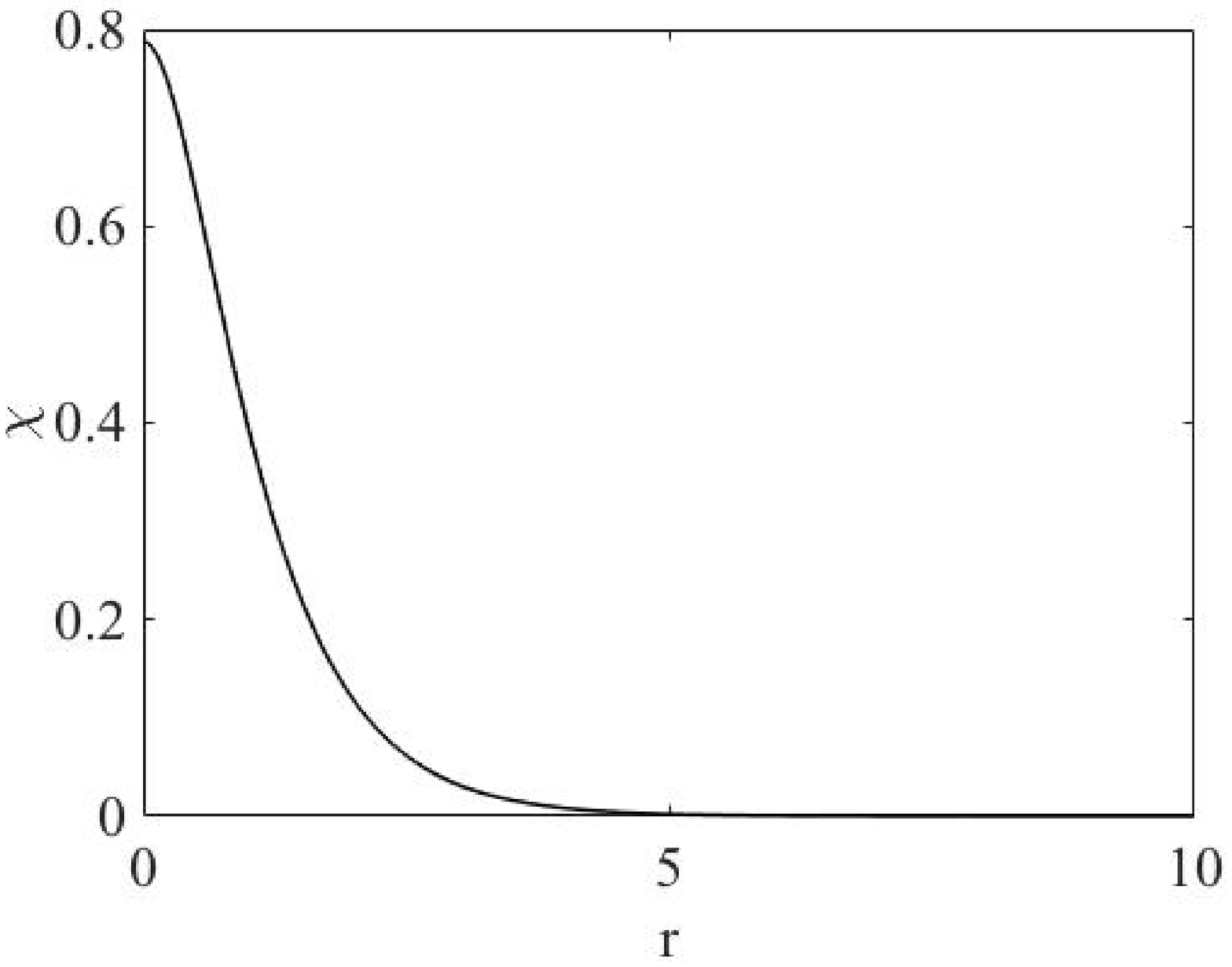}}\subfigure[]{%
\includegraphics[width=3.2in]{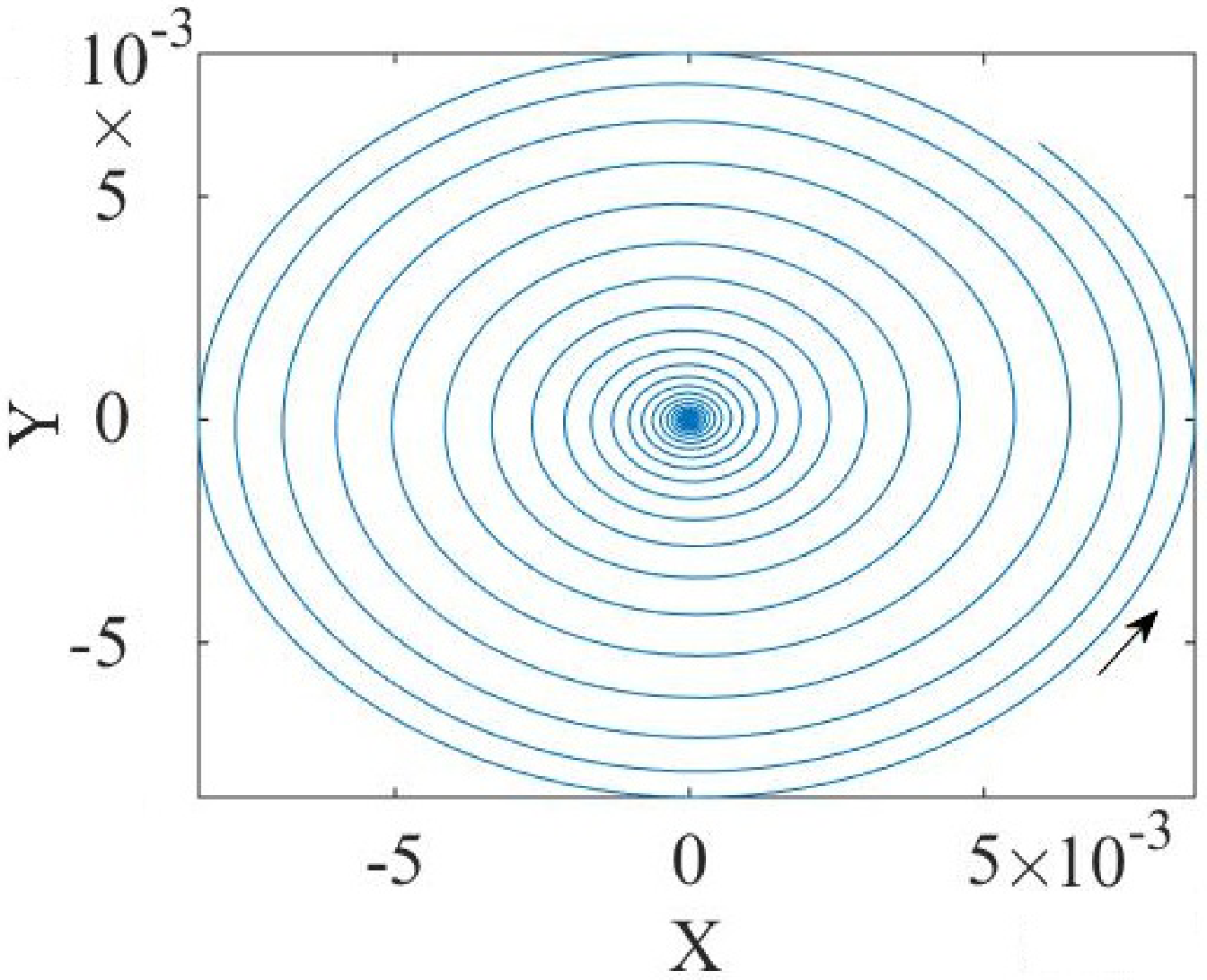}}
\caption{(a) The radial profile of a (weakly unstable) singular vortex mode
with $m=1$, displayed by means of the amplitude function, $\protect\chi (r),$
from which the singularuty is removed by means of transformation (\protect
\ref{chi}), with $U_{0}=1.53$ and $\protect\mu =-1$. The mode's norm is $%
N\approx 4.58$. (b) The instability development of this vortex mode in
direct simulations of Eq. (\protect\ref{varphi}) at $t\leq 7$, is
illustrated by spontaneous motion of its pivot along the spiral trajectory
in the direction indicated by the arrow.}
\label{Fig vortex}
\end{figure}

An example of radial profile $\chi (r)$ of the singular vortex mode, with $%
U_{0}=1.53$, $m=1$ and a finite norm, is displayed in Fig. \ref{Fig vortex}%
(a). This value of $U_{0}$ is chosen in the instability region, close to its
boundary predicted by Eq. (\ref{lstability}), $\left( U_{0}\right) _{\mathrm{%
crit}}^{(1)}\approx \allowbreak 1.56$. In accordance with the prediction
provided by the calculation of eigenmodes of small perturbations, the pivot
of the unstable vortex escapes from the central position, slowly moving away
along a spiral trajectory, as shown in Fig. \ref{Fig vortex}(b).

Eventually, the pivot is ousted to periphery, thus effectively converting
the original unstable vortex into a stable GS with zero vorticity and its
center of mass located at the origin, $x=y=0$. In the course of the
simulations, a large part of the initial norm is consumed by an absorber
installed at the edge of the simulation domain, to emulate losses due to
outward emission of small-amplitude matter waves, in the indefinitely
extended system. In particular, the evolution of the unstable vortex
displayed in Fig. \ref{Fig vortex} leads to its transformation into a
residual GS with norm $N=1.88\approx 41\%$ of the initial value.

The spontaneous transformation of the vortex mode into the GS implies decay
of the mode's angular momentum. In the extended system, the momentum would
be lost with emitted matter waves, while in the present setting it is
gradually eliminated by the edge absorber, as shown by the simulations in
Ref. \cite{Shamriz}.

On the other hand, the simulations demonstrate that perturbed vortices with $%
m=1$ and $m=2$ remain completely stable at, respectively, $U_{0}>14/9$ and $%
U_{0}>77/9$, in agreement with Eq. (\ref{lstability}) \cite{Shamriz}.

\section{Conclusion}

The scenarios for the creation of stable 3D QDs with the help of corrections
to the MF dynamics induced by quantum fluctuations around the MF states (the
LHY effect), proposed in Ref. \cite{Petrov-PRL-2015} and experimentally
realized in binary BEC in Refs. \cite{Cabrera-Sci-2018}-\cite%
{Errico-PRR-2019}, have made a crucially important contribution to the
long-standing problem of the making of stable 2D and 3D soliton-like modes.
A similar mechanism was also experimentally implemented in the
single-component condensate with long-range interactions between atomic
magnetic moments \cite{FB-PRL-2016}-\cite{Chomaz-PRX-2016}. In the
experiment, full 3D QDs \cite{Kartashov-PRA-2018}, as well as oblate
nearly-2D ones \cite{Li-PRA-2018}, have been created as the GS (ground
state), i.e., without embedded vorticity. On the other hand, recent
theoretical analysis has predicted stable vortical QDs, in the fully 3D and
reduced 2D forms, with the unitary and multiple vorticities alike. A very
recent addition to the analysis has demonstrated the existence of stable 2D
vortex modes pulled to the center by the inverse-square potential (\ref{U}),
in which the quantum collapse is suppressed by the LHY effect \cite{Shamriz}%
. These experimental and theoretical results are summarized in the present
review.

There are other directions of the current work on this topic (chiefly,
theoretical ones) which are not included in this brief review, but should be
mentioned: two-component QDs with the linear Rabi mixing between the
components \cite{Cappellaro-2017-SR}, quasi-1D QDs and the spectrum of
excitations in them \cite{Astrakharchik-2018-PRA,Tylutki}, ``quantum balls" stabilized by
the repulsive three-body quintic interaction acting in combination with the LHY correction
and collisions between such ``balls" \cite{Adhikari-2017-PRA,Gautam-2019-JPB},
QDs composed of Bose-Fermi mixtures \cite{Cui-2018-PRA, Desalvo-2017-PRL, Adhikari-2005-PRA, Adhikari-2018-LPL}, QDs in periodic systems \cite{Zhou-2019-CNSNS, Dong-2020-ND, Zheng-2021-FP}, supersolid crystals built of QDs
(see a recent review \cite{Bottcher-2020}), miscibility-immiscibility transitions in dipolar QDs \cite{arXiv1,arXiv2}, and others. Moreover, recent non-perturbative analysis of the strong beyond-mean-field interactions provide extension
of the work to the area where the perturbative LHY model is not valid
\cite{FB-2019-PRR, Staudinger-2018-PRA, Gautam-2019-AP}.

As concerns relevant directions for the continuation of the work, the
creation of the theoretically predicted stable 3D and nearly-2D droplets
with embedded vorticity is a challenging aim. There are also interesting
possibilities for the development of the theoretical analysis, such as the
further consideration of interactions of QDs, as suggested by Ref. \cite%
{Ferioli-PRL-2019}, and a possibility of forming their mutually-orbiting
bound states, precession of 3D droplets with embedded vorticity, considered
as gyroscopes (cf. Ref. \cite{gyro}), the motion of QDs in external
potentials, etc.

\section*{Acknowledgments}

YL acknowledges the supports of the National Natural Science Foundation of
China (Grants Nos. 11874112 and 11905032), the Key Research Projects of General Colleges in Guangdong Province through grant No. 2019KZDXM001, the Foundation for Distinguished Young Talents in Higher Education of Guangdong through grant No. 2018KQNCX279. The work of BAM on this topic is
supported, in part, by grant No. 1286/17 from the Israel Science Foundation.
This author appreciates hospitality of the Department of Applied Physics at
the South China Agricultural University, and collaborations with several
other colleagues on the topics of the present review: G. E. Astrakharchik,
M. Brtka, Z. Chen, R. Driben, A. Cammal, Y. V. Kartashov, K. Kasamatsu, A.
Khare, B. Li, A. Maluckov, T. Meier, T. Mithun, D. S. Petrov, E. Shamriz, Y.
Shnir, L. Tarruell, L. Torner, and M. Tylutki.

\end{document}